\documentclass[aps,twocolumn,superscriptaddress,nofootinbib,a4paper]{revtex4-2}

\usepackage{booktabs}
\usepackage{times}
\usepackage{epsfig}
\usepackage{amsfonts}
\usepackage{amsmath}
\usepackage{amsthm}
\usepackage{amssymb}					
\usepackage{amsthm}
\usepackage{dsfont}
\usepackage{bm}
\usepackage{mathtools}
\usepackage{color}
\usepackage{multirow}
\usepackage[normalem]{ulem}
\newcommand{\stkout}[1]{\ifmmode\text{\sout{\ensuremath{#1}}}\else\sout{#1}\fi}
\usepackage{latexsym}
\usepackage{mathrsfs}
\usepackage{natbib}
\usepackage{verbatim}
\usepackage[T1]{fontenc}
\usepackage{float}
\usepackage{graphicx}
\usepackage{subcaption}
\usepackage{xcolor}
\usepackage{physics}
\usepackage{soul}
\usepackage{caption}
\usepackage{subcaption}

\usepackage{centernot}
\usepackage[font=small,labelfont=bf]{caption}

\usepackage{xspace}

\newtheorem{theorem}{Theorem}

\newtheorem{corollary}{Corollary}

\newtheorem{lemma}{Lemma}

\newtheorem{observation}[theorem]{Observation}
\newtheorem{result}{Result} 

\newtheorem{definition}{Definition}

\newcommand{\floor}[1]{\lfloor #1 \rfloor}

\usepackage{amssymb}
\usepackage{amsthm}
\usepackage{mathtools} 
\usepackage[customcolors]{hf-tikz} 
\usetikzlibrary{patterns}
\usetikzlibrary{matrix,decorations.pathreplacing}

\pgfkeys{tikz/mymatrixenv/.style={decoration={brace},every left delimiter/.style={xshift=8pt},every right delimiter/.style={xshift=-8pt}}}
\pgfkeys{tikz/mymatrix/.style={matrix of math nodes,nodes in empty cells,left delimiter={[},right delimiter={]},inner sep=1pt,outer sep=1.5pt,column sep=2pt,row sep=2pt,nodes={minimum width=20pt,minimum height=10pt,anchor=center,inner sep=0pt,outer sep=0pt}}}
\pgfkeys{tikz/mymatrixbrace/.style={decorate,thick}}

\usepackage{ytableau}
\ytableausetup{boxsize=1.1em}

\usepackage[colorlinks=true,linkcolor=blue,citecolor=magenta,urlcolor=blue]{hyperref}

\begin{document}
	

\title{The classical complexity of quantum state transmission}
	
\author{Elna Svegborn}
\affiliation{Physics Department and NanoLund, Lund University, Box 118, 22100 Lund, Sweden.}
\author{Alexander Bernal}
\affiliation{Physics Department and NanoLund, Lund University, Box 118, 22100 Lund, Sweden.}
\author{Armin Tavakoli}	
\affiliation{Physics Department and NanoLund, Lund University, Box 118, 22100 Lund, Sweden.}
	

\begin{abstract}
Quantum communication provides a natural means to transmit quantum information, but how difficult is it to make such a transmission using only classical communication resources? We address this question by introducing classical communication models for quantum preparations and defining their minimum cost as the \emph{classical complexity}. We prove that this complexity equals the amount of classical communication required to simulate quantum correlations in prepare-and-broadcast experiments, thus establishing a resource equivalence between quantum state transmission and quantum correlations. This calls for a general characterisation of the classical complexity of the noisy quantum state space,  which we show grows exponentially with the quantum dimension and as a power-law in the inverse noise. It reveals the rapidly scaling fundamental cost of classical simulation. Moreover, we develop a toolbox of methods to certify classical complexity in quantum preparation devices, including an exact semidefinite-programming characterisation, analytical witnesses for symmetric preparations and  convex programming scalable up to hundreds and sometimes thousands of qubit states. Our results establish classical complexity as a fundamental measure of quantum communication advantage, revealing the intrinsic cost of classically simulating quantum theory and providing practical tools for its certification.
\end{abstract}

\date{\today}	
\maketitle

\section{Introduction}	
\begin{figure}
	\includegraphics[width=\columnwidth]{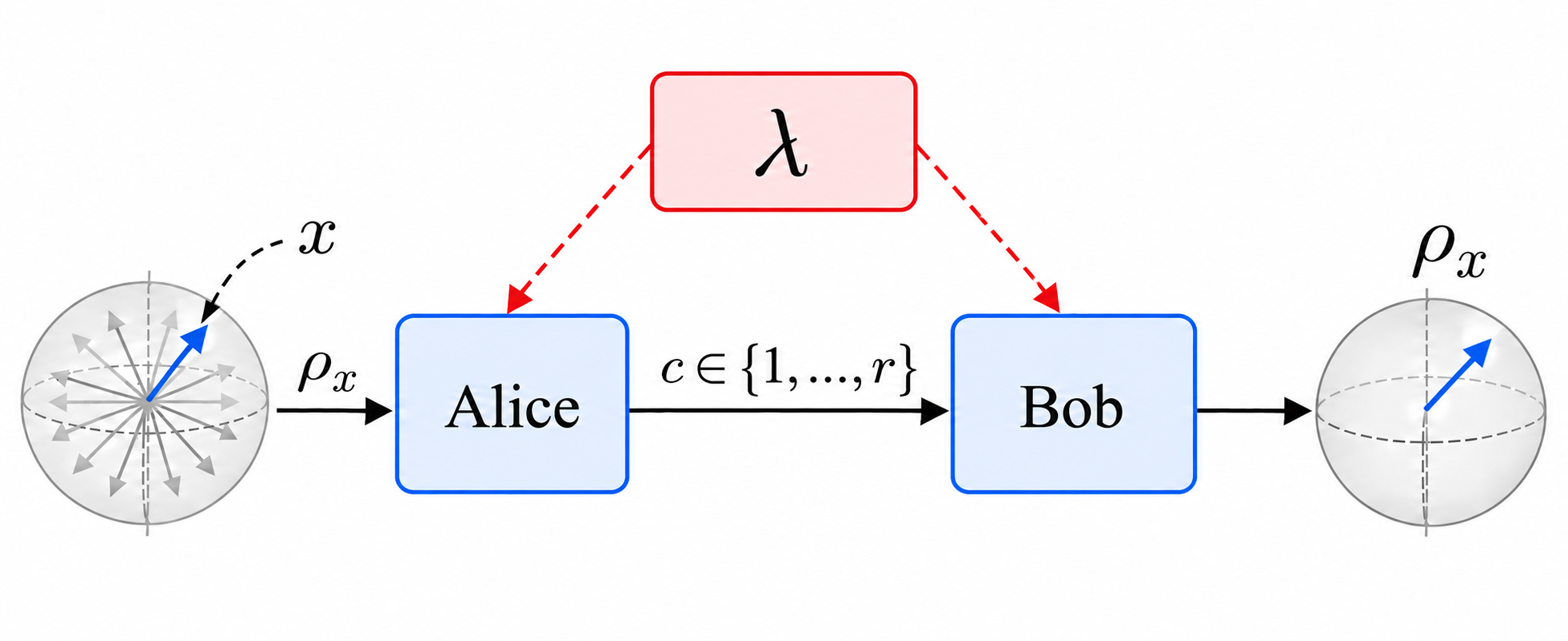}
	\caption{Classical communication model. Alice and Bob are given a classical description of a set of quantum states. Alice selects a state $\rho_x$ and communicates it to Bob using only shared randomness $\lambda$ and a classical message $c$ drawn from an alphabet of $r$ symbols. Bob must generate in his lab  the quantum state $\rho_x$.}\label{fig_scenario}
\end{figure}

Transmitting quantum information between distant locations is a basic primitive of quantum information science. It underpins  applications ranging from quantum cryptography \cite{Gisin2002} to prepare-and-measure protocols \cite{Brask2026, Oliveira2026}. Quantum states can be transmitted directly through a quantum channel or, given shared entanglement, prepared remotely using classical communication \cite{Bennett2001}. Both approaches rely on quantum resources. Understanding the advantage of these quantum communication resources requires quantifying the classical resources needed to accomplish the same task. It leads us to raise the following question.
\begin{quote}
	\textit{How much classical communication is required to transmit a given set of quantum states?}
\end{quote}
Consider two  laboratories that share no quantum resources but have access to shared randomness and a classical communication channel, as illustrated in Fig~\ref{fig_scenario}. The sender, Alice, and receiver, Bob, agree on a predetermined set of quantum states. Alice selects a state $\rho_x$ from this set and communicates enough classical information for Bob to generate it in his lab. We call the minimum size of the classical message alphabet required for this task the \emph{classical complexity}.

Classical complexity benchmarks the advantage of quantum communication by quantifying the classical resources required to reproduce quantum state transmission itself, independently of any subsequent measurement or information-processing task. 
This differs from the long-established study of quantum reductions in communication complexity \cite{Brassard2003, Buhrman2010}, where quantum advantage is defined relative to a specified classical input-output task. The distinction parallels that between entanglement and Bell nonlocality, where the former is a fundamental quantum property and the latter is its observable manifestation in black-box experiments.

Here, we develop a general framework for the classical simulation of quantum state transmission. We first strengthen the operational significance of classical communication models by connecting them to correlations in black-box communication experiments \cite{Brask2026, Oliveira2026}. In the standard prepare-and-measure scenario, the existence of a classical communication model with complexity $r$ guarantees that all correlations obtained by measuring the states can be reproduced using a classical message drawn from an alphabet of $r$ symbols. More importantly, we prove that this condition becomes both necessary and sufficient in the recently introduced prepare-and-broadcast scenario \cite{Sarubi2026}. This establishes an equivalence between the classical resources required to simulate quantum state transmission and those required to reproduce its observable correlations in black-box experiments. This has  both qualitative and quantitative implications.

Qualitatively, this result reveals a sharp distinction between standard prepare-and-measure scenarios and more sophisticated experiments for analysing quantum correlations. In the former, simulation of all $d$-dimensional quantum correlations requires an unbounded classical message alphabet if and only if $d\geq 4$ \cite{Renner2023, Gois2026}, while in the prepare-and-broadcast scenario, the infinite cost arises already for qubits. The distinction is stronger still when communication is quantified by the average number of transmitted bits. That cost is finite in the prepare-and-measure scenario  \cite{Castellano_2026}, but infinite in the prepare-and-broadcast scenario for every $d$. Thus, classical complexity exposes important qualitative features of quantum correlations that remain hidden in standard prepare-and-measure experiments.

Beyond these qualitative distinctions, the resource equivalence strongly motivates the following quantitative question. How large is the classical simulation cost when quantum states are subject to noise? We address this question by considering the entire $d$-dimensional quantum state space subject to isotropic noise. We derive easily computable upper and lower bounds, proving that at any fixed non-unit visibility the required classical complexity grows exponentially with $d$, while at fixed dimension the complexity diverges as a power-law of order $d-1$ in the inverse noise. This confines to a concrete, and rapidly expensive, window the classical cost of simulating not only the noisy quantum state space, but also all quantum correlations that the theory can produce from it. For example, already for $d=16$ and visibility $v=90\%$, the number of classical symbols can be no smaller than $10^{15}$.

While these bounds concern the entire noisy quantum state space, real-life preparation devices typically target a finite set of states. We therefore turn to the practical question of how to certify the classical complexity of such finite sets.  Compatibility with a prescribed classical communication model admits an exact semidefinite programming characterisation, but the size of the program grows rapidly with the number of states, dimension, and classical complexity. We overcome this computational bottleneck using two complementary approaches. For highly symmetric preparations, we derive analytical witness criteria applicable in arbitrary dimension. For qubit preparations, we develop scalable linear-programming methods that yield near-exact characterisations for sets containing several hundreds, and sometimes also many thousands, of states. Together, these methods enable the certification of classical complexity far beyond the regime accessible to standard semidefinite programming \cite{Tavakoli_2024}.

\section{Classical communication models}\label{secCMM}
Consider the scenario illustrated in Fig~\ref{fig_scenario}.  Alice has a classical description of a known set of  quantum states, $\mathcal{E}\equiv \{\rho_1,\rho_2,\ldots,\rho_n\}$. There are $n$ states in the set and their Hilbert space dimension is $d$. She  selects one of the states by choosing a label $x\in\{1,2,\ldots,n\}\equiv [n]$. Her goal is to provide  information to Bob so that he can generate  $\rho_x$.  For this purpose, Alice and Bob share classical communication resources. This consists of a source of shared randomness, $\lambda$, and a classical communication channel supporting messages, $c$, selected from an alphabet of size $r$. Bob uses both $c$ and $\lambda$ to decode the communication into his final quantum output $\rho_x$. Thus, the task is to  reliably communicate any quantum state in $\mathcal{E}$ using at most $r$-valued messages. We call this a classical communication model (CCM).

\begin{definition}[Classical communication model]
	Let $\mathcal{E}=\{\rho_x\}_x$ be a set of $d$-dimensional density matrices. We say that  $\mathcal{E}$ admits an $r$-complex classical communication model if it can be written in the form
	\begin{equation}\label{rcomplex}
		\rho_x = \int d\lambda\ q(\lambda) \sum_{c=1}^r p(c|x,\lambda) \sigma_{c,\lambda},\quad  \forall x,
	\end{equation}
	where $q(\lambda)$ is a probability density,  $p(c|x,\lambda)$ is a conditional  distribution and  $\sigma_{c,\lambda}$ are quantum states. 	The smallest  $r$ for which such a model exists is called the classical complexity of $\mathcal{E}$ and is denoted $r^*(\mathcal{E})$.
\end{definition}
Note that in this definition, the number of states in $\mathcal{E}$ does not have to be finite, which is the reason for the integral in Eq~\eqref{rcomplex}.	Furthermore, we will for simplicity say that $\mathcal{E}$ is \textit{$r$-complex} if it admits the classical communication model \eqref{rcomplex} for the given $r$. The set of all $\mathcal{E}$ that admit an $r$-complex model is denoted $\mathcal{S}_r$.

\subsection{Elementary properties}\label{section_ElementaryProperties}

We begin with listing some elementary properties of CCMs.

\begin{enumerate}
	\item The set $\mathcal{S}_r$ is closed and convex. The sets are also  nested, in the sense that $\mathcal{S}_1\subset \mathcal{S}_2 \subset \mathcal{S}_3\subset \ldots$. 
	
	\item The case of $r=1$ is trivial. That is because $p(c|x,\lambda)$ drops out of Eq~\eqref{rcomplex}, causing the RHS  to lose its dependence on $x$. Hence,  $\mathcal{E}\in \mathcal{S}_1$ if and only if $\mathcal{E}$ contains a single unique state.
	
	\item The membership problem for $\mathcal{S}_r$ is non-trivial only when $n>r$. Otherwise, Alice can simply send the message $c=x$ to Bob, who then knows the classical description of $\rho_x$, and  uses a quantum preparation device to generate it.	
		
	\item In the special case where the CCM is deterministic, i.e.~when there is only one relevant value of  $\lambda$, Eq~\eqref{rcomplex} reduces to $\rho_x=\sum_{c=1}^r p(c|x)\sigma_c$. The classical complexity is then the smallest number of states whose convex hull contains $\mathcal{E}$.	
		
	\item The classical complexity cannot increase by discarding states from $\mathcal{E}$. In other words, $\mathcal{E}'\subset \mathcal{E}$ implies $r^*(\mathcal{E}')\leq r^*(\mathcal{E})$. It also cannot increase by sending all the states in $\mathcal{E}$ through the same quantum channel. Applying a quantum channel, $\Lambda$, to the RHS of \eqref{rcomplex} results in another valid decomposition with $\sigma_{c,\lambda}\rightarrow \Lambda(\sigma_{c,\lambda})$.
	
	\item If we were to extend a set $\mathcal{E}$ with every state obtained from taking convex combinations of its members, the classical complexity would remain unchanged. In other words, $r^*\left(\text{conv }\mathcal{E}\right) = 	r^*\left(\mathcal{E}\right)$. This shows that the size of $\mathcal{E}$ can be extended indefinitely while the  classical complexity remains constant. 
	
	\item If $\mathcal{E}$ contains only pure states, we must have $\rho_x=\sigma_{c,\lambda}$ whenever $q(\lambda)p(c|x,\lambda)>0$. Thus, we can w.l.g.  ~choose $\lambda$ deterministic. Moreover, if all pure  states are distinct, $\mathcal{E}$ is $r$-complex if and only if $r\geq n$. Hence, the classical complexity is $r^*(\mathcal{E})=n$. We conclude that for pure quantum states, the classical complexity is unbounded as the number of states increases.
	
\end{enumerate}

\subsection{Semidefinite programming characterisation}

A central question is whether a given set of quantum states is compatible with an $r$-complex model. Conveniently, a few elementary observations reveal a  method of answering this question. To this end, consider that the number of states, $n$, is fixed. We will express the response functions, $p(c|x,\lambda)$, as convex combinations of deterministic distributions. To this end, let $D_\mu(c|x)=\delta_{f_\mu(x),c}$ be the deterministic distribution associated with every function $f$ that maps $x$ into $c$. As there are $r^n$ such functions, they are here indexed by $\mu=1,\ldots,r^n$.  We can therefore write $p(c|x,\lambda)=\sum_\mu p(\mu|\lambda) D_\mu(c|x)$. Inserting this in Eq~\eqref{rcomplex}  and defining  $p(\mu,\lambda)=q(\lambda)p(\mu|\lambda)$ gives $\rho_x = \sum_{\mu,c} D_{\mu}(c|x) \int d\lambda\  p(\mu,\lambda)\sigma_{c,\lambda}$. Note that $\tilde{\sigma}_{c,\mu}\equiv \int d\lambda\  p(\mu,\lambda)\sigma_{c,\lambda}$ represents an arbitrary convex combination of states, but with a fixed sub-normalisation $p(\mu)$. This leads us to our first result.

\begin{observation}[Semidefinite characterisation]\label{Result1}
Let $\mathcal{E}=\{\rho_x\}_{x=1}^n$ be a finite set of quantum states. It admits an $r$-complex model if and only if the following semidefinite program is feasible
\begin{equation}\label{SDPrcomplex}
\begin{aligned}
\text{find}\quad
& \{\tilde{\sigma}_{c,\mu}\}_{c,\mu}  \\
\text{s.t.}\quad
& \rho_x
= \sum_{\mu=1}^{r^n}\sum_{c=1}^{r}
D_\mu(c|x)\tilde{\sigma}_{c,\mu},
&& \forall x, \\
& \tr(\tilde{\sigma}_{c,\mu})=\tr(\tilde{\sigma}_{1,\mu}),
&& \forall c,\mu, \\
& \tilde{\sigma}_{c,\mu}\succeq 0,
&& \forall c,\mu .
\end{aligned}
\end{equation}
\end{observation}
This provides a simple and standard tool for determining the existence of an $r$-complex model. In particular, running this program separately for $r=1,2,\ldots,r^*(\mathcal{E})$ (i.e.~increasing $r$ until it becomes feasible) also gives us a complete criterion for the classical complexity of $\mathcal{E}$. Moreover, by considering the dual formulation of the SDP \eqref{SDPrcomplex} we can obtain witness criteria for falsifying $r$-complex models, as discussed in Appendix~\ref{AppPrimalDual}. However, due to the exponential scaling of $\mu$, these SDPs become computationally expensive quite soon beyond the few smallest choices of $(n,d,r)$.  In Appendix~\ref{AppSterlingStrategies}, we show that the program can be  simplified w.l.g.  by restricting the alphabet of $\mu$, gaining a variable reduction of roughly a factor $r!$, but this only moderately extends practical computation in the parameter $r$.

\subsection{Relation to other notions of classicality}
We contextualise our proposed notion of CCMs by discussing its relation to other, already established, notions of classicality. An important distinction is that CCMs define classicality directly at the level of quantum states, whereas established  models assess classicality at the level of correlations \cite{Buhrman2010, Brask2026, Oliveira2026}. Consider first the conventional prepare-and-measure (PM) scenario (see e.g.~the review \cite{Brask2026}). Alice receives an input $x$ and sends Bob a $d$-dimensional quantum state $\rho_x$. Bob chooses an input $y$ and performs a POVM $\{M_{b|y}\}_b$, producing correlations
\begin{equation}\label{born}
p(b|x,y)=\tr(\rho_xM_{b|y}).
\end{equation}
These correlations admit a classical simulation with an $r$-valued message if \cite{Gallego2010}
\begin{equation}\label{PMclassicalmodel}
p(b|x,y)=\int d\lambda \ q(\lambda)\sum_{c=1}^r p(c|x,\lambda)p(b|y,c,\lambda).
\end{equation}
The case $r=d$ is most commonly considered in the literature because it corresponds to comparing classical and quantum messages of the same dimension.

Given a set of states $\mathcal{E}=\{\rho_x\}_x$, determining whether every correlation obtained from it admits the model \eqref{PMclassicalmodel} is generally difficult. Previous approaches have focused on qubits, projective measurements, and $r=2$ \cite{Gois2021}. CCMs provide a simple sufficient criterion for this problem, that applies without any of these restrictions.
\begin{observation}\label{Obs1}
	Let $\mathcal{E}=\{\rho_x\}_x$ be a set of $d$-dimensional quantum states. If $\mathcal{E}\in\mathcal{S}_r$, then for every set of POVMs the resulting correlations \eqref{born} admit the decomposition in Eq~\eqref{PMclassicalmodel}.
\end{observation}
\begin{proof}
	If $\mathcal{E}$ admits an $r$-complex model, then we obtain directly from Eq~\eqref{rcomplex} and Eq~\eqref{born} that $p(b|x,y)=\int d\lambda q(\lambda)\sum_{c=1}^r p(c|x,\lambda)\tr(\sigma_{c,\lambda}M_{b|y})$. We recover Eq~\eqref{PMclassicalmodel} by identifying $p(b|y,c,\lambda)=\tr(\sigma_{c,\lambda}M_{b|y})$.
\end{proof}
Thus, an $r$-complex CCM implies that $\mathcal{E}$ cannot generate correlations in the PM scenario that elude modelling with an $r$-valued classical message. When combined with Observation~\ref{Result1}, this provides a computable criterion for establishing quantum states that have only classical correlations in the PM scenario. 

However, this criterion is not complete, because a set of states may generate only classical PM correlations without admitting the corresponding CCM. This separation parallels that between entanglement and Bell nonlocality, where entangled states need not generate nonlocal correlations \cite{Werner1989}. An extreme example of this is obtained from the Frenkel--Weiner theorem \cite{Frenkel2015}, which shows that the quantum correlations obtained from making a fixed measurement on a $d$-dimensional system can be simulated classically with messages of the same dimension. In contrast, the classical complexity of such  a set of states can in principle be unbounded (see Property 7 in section~\ref{section_ElementaryProperties}). Most importantly, as we will show in the next section, the separation disappears when venturing beyond the PM scenario, to a setting that can reveal the full power of such quantum correlations without introducing extra assumptions.

We can also compare CCMs with the incoherent preparation models introduced in Refs~\cite{Cobucci2026, Cobucci2026b}. Like our CCMs, these models concern the states themselves, but their focus is not on transmitting quantum states but instead on whether they can be generated without superposition. One can view these incoherent state models as CCMs subject to additional constraints, specifically  $r\leq d$ and $\sigma_{c,\lambda}=U_\lambda\ketbra{c}{c}U_\lambda^\dagger$ for some unitary $U_\lambda$. CCMs impose neither of these two restrictions, since Bob may prepare arbitrary states $\sigma_{c,\lambda}$ and the models remain well-defined for $r>d$. In particular, for $r=d$, incoherent models form a strict subset of CCMs, and the distinction is significant. For example, two noncommuting pure qubit states do not admit an incoherent model because they are in relative superposition, but they trivially admit a $2$-complex CCM.

\section{Operational meaning of classical communication models}\label{Sec:operational}
Observation~\ref{Obs1} establishes that any set of states admitting an $r$-complex CCM can generate only classical correlations in a PM experiment, but the converse does not hold. The latter prevents $r$-complexity from becoming the fundamental resource for quantum correlations in the PM scenario. However, as we show here, this is a limitation not of the concept of CCMs but of restricting the analysis of correlations to the PM scenario. In this section, we show that in a more sophisticated experiment, called the  prepare-and-broadcast (PAB) scenario \cite{Sarubi2026}, $r$-complex CCMs are not only sufficient for classical correlations but also necessary. This provides an operational interpretation that bridges the notion of quantum advantage in transmitting states with that of generating quantum correlations.

\begin{result}[Operational interpretation]\label{Result2}
A set of quantum states can generate correlations in a prepare-and-broadcast experiment that cannot be simulated using an $r$-valued classical message if and only if the set admits no $r$-complex classical communication model.
\end{result}

\begin{figure}[t!]
	\includegraphics[width=0.7\columnwidth]{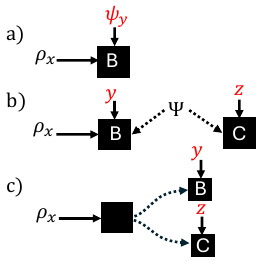}
	\caption{Communication scenarios. \textbf{a)} MDI scenario, where $B$ has quantum inputs $\psi_y$. \textbf{b)} DI extension of MDI scenario. \textbf{c)}  PAB scenario.  Classical outputs have been omitted for simplicity.}\label{fig_scenario_MDI_DI_PB}
\end{figure}

To understand the meaning of this result, let us first introduce the PAB scenario. Alice's input selects a $d$-dimensional quantum state from a set $\mathcal E = \{\rho_x\}_x$, but in contrast to the PM scenario they are then passed through a broadcasting device that implements some arbitrary quantum transformation. This completely positive trace-preserving map, $\Lambda$, creates a bipartite state which is distributed to two receivers, $B$ and  $C$, who perform measurements selected by their respective inputs $y$ and $z$; see  Fig~\ref{fig_scenario_MDI_DI_PB}c). Thus, the quantum correlations become
\begin{equation}\label{QPAB}
	p(\tilde{b},\tilde{c}|x,y,z)=\tr\left(\Lambda[\rho_x]M^{B}_{\tilde{b}|y}\otimes M^{C}_{\tilde{c}|z}\right),
\end{equation}
where $\tilde{b}$ and $\tilde{c}$ are the outputs of $B$ and $C$ respectively. In contrast, correlations compatible with  Alice sending $r$-valued classical messages   take the form \cite{Sarubi2026}
\begin{equation}\label{PABclassical}
	p(\tilde{b},\tilde{c}|x,y,z)=\int d\lambda\, q(\lambda) \sum_{c=1}^r p(c|x,\lambda) p_Q(\tilde{b},\tilde{c}|c,y,z,\lambda),
\end{equation}
where $p_Q$ is an arbitrary quantum probability distribution compatible with the no-signalling principle (see also \cite{Bowles2021}). Such a general $p_Q$ is the natural choice, because we want to ensure that the quantum features are due to Alice's preparations and not generated by the broadcasting device.

Result~\ref{Result2} can therefore be understood as two separate statements. First, if Alice's states admit an $r$-complex model, then all quantum distributions $p(\tilde{b},\tilde{c}|x,y,z)$ achievable in the PAB scenario admit the  model \eqref{PABclassical}. This is a simple  extension of Observation~\ref{Obs1}: insert Eq~\eqref{rcomplex} in Eq~\eqref{QPAB} and define $\tr \small(\Lambda[\sigma_{c,\lambda}]M_{\tilde{b}|y}\otimes M_{\tilde{c}|z}\small)=p_Q(\tilde{b},\tilde{c}|c,y,z,\lambda)$ to recover Eq~\eqref{PABclassical}. Second, and more interestingly, if Alice's states do not admit an $r$-complex model, then there exists a quantum distribution $p(\tilde{b},\tilde{c}|x,y,z)$ that does not admit the model in Eq~\eqref{PABclassical}. In this sense, Result~\ref{Result2} tells us that \textit{$r$-complex CCMs are the defining feature of quantum correlations with respect to classical theories of any chosen complexity}. The rest of this section is devoted to proving this result. 

We provide an overview of our proof. Inspired by Refs~\cite{Buscemi_2012, Branciard_2013}, we will begin with a PM scenario modified so that party $B$'s classical input alphabet is replaced with a tomographically complete set of quantum states; see Fig~\ref{fig_scenario_MDI_DI_PB}a). In this setting, we prove a one-to-one mapping between $r$-complexity and the corresponding measurement-device-independent (MDI) classical correlations. Next, we remove the need for having the quantum inputs for $B$.  This is achieved by introducing an additional measuring party, $C$, who shares an entangled state with  $B$; see Fig~\ref{fig_scenario_MDI_DI_PB}b). By employing device-independent self-testing techniques \cite{Bowles_2018_pra, Bowles_2018_prl} on the marginal distribution between $B$ and $C$, we self-test that $C$ remotely prepares an informationally complete set of sub-normalised states for $B$. Since this set can be interpreted as the input states in the MDI scenario, this allows us to promote the MDI proof into a fully DI witness. Finally, we exploit the observation of Ref~\cite{Sekatski2025}, namely that the extra source can be incorporated into the broadcasting device of the PAB scenario, thus yielding Result~\ref{Result1}.

\subsection{Measurement-device-independent certification}\label{secMDI}
Consider that Alice privately selects a classical input $x$ and encodes it into a state $\rho_x$ sent to Bob. This communication can either be a $d$-dimensional quantum message or an $r$-valued classical message. Upon receiving Alice's message, Bob selects a quantum input from an informationally complete set of states $\{\psi_y\}_y$, and then performs a joint measurement $\{B_b\}_b$ on the two systems with  outcome $b$. The probabilities are given by Born's rule
\begin{equation}\label{eq:MDI_corr}
	p(b|x,\psi_y) = \tr(\psi_y  M_{b|x}),
\end{equation}
where we have defined the effective POVM $M_{b|x} = \tr_A(\left[\rho_x \otimes \openone\right]B_b)$ acting on $\psi_y$.

When communication is classical, Alice uses a response $p(c|x,\lambda)$ to encode $x$ into $c\in\{1,\ldots,r\}$ using the shared randomness $\lambda$. Bob reads $c$ and uses it together with  $\lambda$ to choose a quantum measurement $\{G_{b|c,\lambda}\}_b$ acting on his quantum input state $\psi_y$. Using Eq~\eqref{eq:MDI_corr}, these classical MDI correlations can equivalently be expressed as 
\begin{equation} \label{eq:M_classical}
M_{b|x} = \int d\lambda\, q(\lambda) \sum_{c=1}^r p(c|x,\lambda) G_{b|c,\lambda}, \quad \forall b,x
\end{equation}
By definition, every $r$-complex set  $\{\rho_x\}_x$ satisfies Eq~\eqref{eq:M_classical}. 

Due to convexity, any $\mathcal M = \{M_{b|x}\}_{b,x}$ that does not admit the decomposition  \eqref{eq:M_classical} can be detected as such through a separating hyperplane. This corresponds to a witness $\mathcal W_r(\mathcal M) = \sum_{b,x} \tr(\mathcal O_{bx} M_{b|x}) \leq 0$, where $\mathcal{O}_{bx}$ are appropriate Hermitian operators and the bound applies to all $\mathcal{M}$ satisfying \eqref{eq:M_classical}. Expand $\mathcal O_{bx}$ in the informationally complete set $\{\psi_y\}_y$, namely $\mathcal O_{bx} = \sum_{y} a_{bxy} \psi_y$ for some real coefficients  $a_{bxy}$. Hence, the witness together with Eq~\eqref{eq:MDI_corr} provides an MDI inequality
\begin{equation}\label{eq:witnessmdi}
\mathcal{W}_r(\mathcal M) = \sum_{b,x,y} a_{bxy} p(b|x,\psi_y) \leq 0.
\end{equation}
We now construct an explicit witness for which a violation is possible if and only if $\mathcal E \notin \mathcal S_r$.

Let the measurement  $\{B_b\}_b$ be a projection onto an orthonormal basis of maximally entangled states, $\ket{\phi_b} = (\openone \otimes U_b)\ket{\phi^+}$, where $\ket{\phi^+} = \sum_{i = 0}^{d-1} \ket{i,i}/\sqrt{d}$ is the maximally entangled state and $\{U_b\}_{b=1}^{d^2}$ is a unitary basis. This gives $M_{b|x}=\frac{1}{d}U_b \rho_x^T U_b^\dagger$. If  $\mathcal M$ admits the form  \eqref{eq:M_classical}, then $\frac{1}{d}U_b \rho_x^T U_b^\dagger = \int d \lambda q(\lambda)\, \sum_{c=1}^r p(c|x,\lambda) G_{b|c,\lambda}$. Solving this equation for the preparations $\rho_x$, and summing both sides over all possible outcomes $b$ gives $\rho_x = \int d \lambda \, q(\lambda) \sum_c p(c|x,\lambda) \tau_{c,\lambda}$, where we have defined $\tau_{c,\lambda}  \equiv \frac{1}{d} \sum_b U_b^T G_{b|c,\lambda}^T U_b^*$. Since $\tau_{c,\lambda}$ is a state, we have recovered the definition of $\{\rho_x\}_x$ being $r$-complex. Thus, the decomposition \eqref{eq:M_classical} implies an $r$-complex CCM. By the contrapositive, the failure of an $r$-complex CCM implies the failure of the model \eqref{eq:M_classical}.

 \subsection{Device-independent certification}\label{sec:DI_cert}
 We now remove the trust assumption associated with the quantum inputs in the MDI scenario by using arguments from self-testing. To this end, consider the  scenario in Fig~\ref{fig_scenario_MDI_DI_PB}b), where we have added a source of entanglement, $\Psi$, and an extra party $C$. The inputs of the three parties are $x$, $y$ and $z$, respectively, but only $B$ and $C$ have outputs, labeled by $\tilde b$ and $\tilde c$. The quantum correlations in this scenario become
 \begin{equation}\label{eq:correlations_DI}
 	p(\tilde b, \tilde c|x,y,z) = \tr[\left(\rho^A_x \otimes \Psi^{BC}\right)\left(B_{\tilde b|y}^{AB} \otimes C_{\tilde c|z}^C\right)].
 \end{equation}
  Here, $\{B_{\tilde b|y}\}_{\tilde b}$ and $\{C_{\tilde c|z}\}_{\tilde c}$ are the POVMs of the two measuring parties. Since the spaces of $B$ and $C$ can have any dimension, we can w.l.g. ~take $\Psi$ pure and the measurements acting on these spaces projective. Contrasting this, a classical model with an $r$-valued message from Alice yields correlations of the form \eqref{PABclassical} with  $p_Q(\tilde b, \tilde c |c,y,z,\lambda) = \tr\small((\sigma_{c,\lambda} \otimes \Psi)(B_{\tilde b|y} \otimes C_{\tilde c|z})\small)$. We now show that there exists a protocol such that if Alice's states do not admit an $r$-complex model, the associated distribution $p(\tilde b, \tilde c|x,y,z)$ is not compatible with any set of correlations achievable with an $r$-valued message. For simplicity, we start with the case of qubit preparations.

 Choose input alphabets $y\in\{1,2,3,4,5,6,\star\}$ and  $z \in \{1,2,3\}$. The outcomes are binary,  $\tilde b, \tilde c \in\{\pm1\}$, except when $y = \star$ for which we have outcomes $\tilde b \in\{ 1,\ldots,d^2\}$. When Bob selects $y = \star$ he performs a joint measurement on the incoming state $\rho_x$ and his share of the entangled state $\Psi$. Otherwise, he only measures on his share of $\Psi$. If for every $x$ the distribution $p(\tilde b, \tilde c|x,y\neq \star,z)$  maximally violates a CHSH-type of Bell test, this self-tests a maximally entangled two-qubit $\Psi$  (up to local isometry) and similarly measurements for $C$ that are  Pauli observables $(\sigma_X,\pm \sigma_Y,\sigma_Z)$ \cite{Bowles_2018_pra,Bowles_2018_prl}. By extension, this self-test provides a device-independent certificate that Bob's assemblage corresponds to the complete set of sub-normalised Pauli measurements (up to possible transposition).

Thus, all settings but $y=\star$ are used to certify the states remotely  prepared for $B$. We then use the setting $y = \star$ for the test of the MDI witness. Specifically, the effective measurements  in section~\ref{secMDI} become $M_{\tilde b|x} \equiv \tr_A([\rho_x \otimes \openone]B_{\tilde b|\star})$. We can then rely on our previous MDI proof to construct a witness inequality
\begin{equation}\label{eq:DI_inequality}
	\mathcal W_r = \sum_{\tilde b,\tilde c,x,z}a_{\tilde b \tilde c xz}p(\tilde b,\tilde c|x,y = \star,z)  \leq 0,
\end{equation}
where the coefficients $a_{\tilde b \tilde cxz}$ are used to decompose the Hermitian witness operators $\mathcal O_{\tilde b,x}$ over the self-tested Pauli eigenstates. Hence, after self-testing, this is violated if and only if the states defy an $r$-complex CCM. It provides us with a device-independent certificate. The proof is given in Appendix~\ref{AppDI} and it is based on ideas that closely parallel those used in the proof of the MDI inequality, replacing $y$ with the pair $(\tilde c,z)$ and indexing Bob's quantum inputs by $\psi_{\tilde c,z}=\frac{1}{d}\pi_{\tilde c|z}$ where $\pi_{\tilde c|z}= \frac{1}{2}(\openone + \tilde c \sigma_z)$ are the eigenstates of the Pauli observables.

This self-testing procedure can be extended beyond qubits.  It is achieved by applying the qubit self-testing protocol many times in parallel to certify that Bob's input states correspond to tensor products of Pauli observables \cite{Bowles_2018_pra} which is sufficient to obtain the required informationally complete set. We refer to Appendix~\ref{AppDI_high_dim} for details. This result parallels that of how every entangled state can be device-independently certified in a Bell scenario  with auxiliary sources of entanglement  \cite{Bowles_2018_prl}.

 \subsection{Prepare and broadcast scenario}
Lastly, we lift the device-independent interpretation of  $r$-complexity to the PAB scenario, illustrated in Fig~\ref{fig_scenario_MDI_DI_PB}c). This is done by integrating the auxiliary source into the broadcasting channel and considering the exact same measurement settings for $B$ and $C$ and self-testing techniques as in section \ref{sec:DI_cert}.  This is, choose a broadcasting channel, $\Lambda$, that relays Alice's quantum state to $B$ and additionally generates a maximally entangled state shared between $B$ and $C$. The self-testing applies to this auxiliary state:  if for every $x$ the distribution $p(\tilde b,\tilde c|x,y \neq \star, z)$ in Eq~\eqref{QPAB} maximally violates a CHSH-type inequality, this provides a device-independent certificate that $\Lambda$ has indeed distributed a maximally entangled state to $B$ and $C$, i.e.~$\Lambda[\rho_x] = \tilde \Lambda_{B'}[\rho_x] \otimes \ketbra{\phi^+}_{BC}$, and that $B$ receives from $C$ an informationally complete set of quantum inputs proportional to tensor products of Pauli eigenprojectors \cite{Sekatski2025}.

Similarly to section \ref{sec:DI_cert}, we then use $B$'s setting $y = \star$ to construct a witness inequality of the form of Eq~\eqref{eq:DI_inequality}, i.e., $\mathcal W_r[p(\tilde b, \tilde c|x,y = \star,z)] \leq 0$. The construction is the previous MDI proof, with the effective measurements $M_{\tilde b|x} = \tr_{B}([\tilde \Lambda(\rho_x) \otimes \openone] B_{b|\star})$.  The inequality is satisfied by all correlations of the form in Eq~\eqref{PABclassical}, but can be violated if Alice's set of preparations does not admit an $r$-complex CCM and the quantum channel e.g. takes the form $\tilde \Lambda(\rho_x) = V \rho_x V^\dagger$, meaning that it simply routes Alice's preparation to Bob up to a unitary rotation $V$. We refer to Appendix~\ref{AppDI} for details. That is, succesful self-test together with $\mathcal W_r > 0$ excludes any $r$-complex CCM. This protocol generalises directly to arbitrary dimension by extending the self-test from qubits to qudits, as mentioned in section~\ref{sec:DI_cert}. The conclusion is Result~\ref{Result2}.

\section{Classical complexity of noisy quantum theory}
Result~\ref{Result2} has shown us that classical complexity determines the communication threshold for generating quantum correlation advantages. This means that the ability to classically reproduce quantum states is ultimately equivalent to the ability to extract black-box quantum correlation advantage from them. This connection provides strong motivation for asking the following question: \textit{what is the classical complexity of the space of all states in noisy quantum theory?} 

Already an elementary observation reveals that the answer ought to be insightful. In the PM scenario, all correlations generated by measuring qubits can be classically reproduced at the mere cost of two bits of communication \cite{Renner2023}. However, the classical complexity of all pure quantum states is infinite (see Property 7 in section \ref{section_ElementaryProperties}). Thus, it follows from Result~\ref{Result2} that by shifting from the PM experiment to  the PAB experiment, we can, at least in principle, unlock unbounded quantum correlation advantage from sending single qubits.

In order to address the classical complexity of noisy quantum theory, we consider  all pure states in dimension $d$ when subject to an isotropic visibility $v\in[0,1]$. This noisy quantum state space is written 
\begin{equation}\label{eq:noisy_quantum_theory}
\mathcal{E}_Q = \left\{\rho_\psi\right\}_{\psi \in \mathbb{C}^d} \quad \text{where} \quad \rho_\psi = v \ketbra{\psi} + \frac{1-v}{d}\openone.
\end{equation}
Our goal is to determine the classical complexity of $\mathcal{E}_Q$, but first we highlight two  trivial extremes of the problem. For noiseless states ($v=1$) the classical complexity is infinite, whereas for fully noisy states ($v=0$) it is unit (see Property 2 in section \ref{section_ElementaryProperties}). Our task is therefore to gain insight on the critical visibility, $v^*(\mathcal{E}_Q)$, below which $\mathcal{E}_Q$ has an $r$-complex model and above which it does not. This is addressed by our next result.
 
\begin{result}[Classical complexity of isotropic quantum theory]\label{ResultCriticalVisibility}
Consider the set of all pure $d$-dimensional states subject to isotropic noise, as given in Eq~\eqref{eq:noisy_quantum_theory}. For $r \geq d $, the critical visibility for an $r$-complex CCM is bounded from below as
\begin{equation}\label{eq:lower_bound}
v^*(\mathcal{E}_Q)\geq  \max\left\{\frac{H_d-1}{d-1}, 1-\prod_{c=2}^r \frac{c}{\frac{1}{d-1}+c}\right\},
\end{equation}
where $H_d = \sum_{k=1}^d \frac{1}{k}$. It is bounded from above as 
\begin{equation}\label{eq:upper_bound}
v^*(\mathcal{E}_Q) \leq  1-r^{-\frac{1}{d-1}}.
\end{equation}
\end{result}

\begin{proof}
We present the main idea used in the proof and refer to Appendix~\ref{AppClassicalModels} for details. The proof is based on a witness test and an explicit $r$-complex CCM of the form of Eq~\eqref{rcomplex}. Thus, take the witness to be $W = \int d \psi \langle \psi |\rho_\psi |\psi\rangle$, i.e., the average overlap between the noisy state $\rho_\psi$ and its associated target state $\psi \in \mathbb C^d$. Given that the set $\{\rho_\psi\}_\psi$ admits an $r$-complex CCM the witness $W$ can at most reach the value
\begin{equation}\label{eq:integral_main}
	\mathcal{I}_{d,r} = \max_{\phi_1,\dots,\phi_r \in \mathbb{C}^d} \int d \psi \max_{1\leq c\leq r} \left|\braket{\psi}{\phi_c}\right|^2.
\end{equation}
We now show  that this value is achievable through the construction of an explicit CCM \eqref{rcomplex}. Denote by $\{\ket{\phi_c^*}\}_c$ the optimal set of $r$ states in Eq~\eqref{eq:integral_main}. Let the CCM use unbounded shared randomness, where each shared variable $\lambda$ is associated to a unique unitary $U$. For fixed $\psi$, we then simulate each $\rho_\psi$ by averaging the state $\tau_{\psi,U}$ over $U\in \mathrm{SU}(d)$, where $\tau_{\psi,U}$ is chosen to be the element in the set of pure states $\{U \ket{\phi_c^*}\}_{c=1}^r$ that has the largest overlap with $\ket \psi$. Identifying the integration over unitary transformations with Haar random pure states $\psi \in \mathbb C^d$, and computing the overlap between $\rho_\psi$ and $\psi$, we recover the integral $\mathcal{I}_{d,r}$ in Eq~\eqref{eq:integral_main}. Thus,  the integral gives a necessary and sufficient condition for $r$-complex CCM. That is, by computing this integral, we can determine the exact value of the visibility $v$ at which the set of noisy states in Eq~\eqref{eq:noisy_quantum_theory} is $r$-complex simulable.  Specifically, this noise threshold is given by $v \leq \frac{d\mathcal{I}_{d,r}-1}{d-1}$. While we cannot solve this integral exactly, we can derive the upper and lower bounds in Eqs~\eqref{eq:upper_bound} and \eqref{eq:lower_bound}, as shown in Appendix~\ref{AppClassicalModels}.
\end{proof}

We make three remarks. First, by property 6 in section~\ref{section_ElementaryProperties}, the lower bound in Result~\ref{ResultCriticalVisibility} applies not only to the noisy states in Eq~\eqref{eq:noisy_quantum_theory} but also to any other set of states contained in $\text{conv } \mathcal{E}_Q$. Second, the proof reduces the characterisation of the critical visibility $v^*\left(\mathcal{E}_Q\right)$  to evaluating the integral \eqref{eq:integral_main}. Only because we are unable to evaluate this analytically have we  focused on the upper and lower bounds appearing in Result~\ref{ResultCriticalVisibility}. Third, for $r\leq d$ the states $\phi_c$ in \eqref{eq:integral_main} are optimally taken to be orthogonal (see Appendix~\ref{AppClassicalModels}). Thus, the explicit construction described above coincides with the incoherent simulation model considered in Result 1 in Ref~\cite{Cobucci2026}. This means that the exact critical visibility for $r \leq d$ is $v^*(\mathcal{E}_Q) = \frac{H_r-1}{d-1}$ \cite{Cobucci2026}.

However, the interesting regime for our purposes is $r\geq d$ and especially $r\gg d$. In  Eq~\eqref{eq:lower_bound}, the first bound is exact for $r=d$ but independent of $r$, whereas the second is not tight but grows with  $r$. Therefore, as we increase $r$ for fixed $d$, the latter will become the relevant one. For large $r$, the lower bound on the visibility grows as  $1-C_d r^{-1/(d-1)}$ with $C_d = \frac{d}{d-1}\Gamma(1+\frac{1}{d-1})$ and $\Gamma$ is the gamma function. Comparing this with the upper bound \eqref{eq:upper_bound}, we see that both scale as $r^{-1/(d-1)}$. Hence, this determines the large-$r$ scaling of $v^*\left(\mathcal{E}_Q\right)$.   Re-arranging this equation, the classical complexity of noisy quantum theory scales as  
\begin{equation}
r^*\left(\mathcal{E}_Q\right) \sim \left(\frac{A_d}{1-v}\right)^{d-1},
\end{equation}
for some dimension-dependent constant $1 \leq A_d \leq C_d$. For large $d$, we have $C_d \rightarrow 1$, see Appendix~\ref{AppClassicalModels}. Hence, the complexity grows exponentially with $d$ at fixed $v$. For a fixed $d$, the complexity grows in $1/(1-v)$ as a power-law with degree $d-1$.

This shows that the classical cost of simulating quantum communication becomes very high both as we increase the dimension or as we increase the visibility. Thus, while under this kind of noise only pure quantum states have infinite classical complexity,  a reasonable dimension and reasonable visibility are, at least in the fundamental sense, sufficient to generate very large separations between quantum and classical models. This is illustrated for a few examples in Table~\ref{tab:bounds}. \\

\begin{table}[t!]
	\centering
	\caption{Lower and upper bounds on classical complexity for different values of $v$ and $d$.}
	\label{tab:bounds}
	\begin{tabular}{c | c c c c}
		\toprule
		$v \backslash d$ & $2$ & $4$ & $16$ & $32$\\
		\midrule
		$0.5$  & $2$--$4$  & $8$--$14$  	& $(3.3$--$5.1)\times 10^{4}$  & $(2.1$--$3.3 )\times 10^{9}$\\
		
		$0.9$  & $10$--$20$ 	& $(1.0$--$1.7) \times 10^{3}$ 	& $(1.0$--$1.6)\times 10^{15}$ & $(1.0$--$1.5)\times 10^{31}$ \\
		
		$0.99$ & $100$--$200$	& $(1.0$--$1.7)\times 10^{6}$	& $(1.0$--$1.6)\times 10^{30}$ & $(1.0$--$1.5 )\times 10^{62}$ \\
		\bottomrule
	\end{tabular}
\end{table}

\textbf{Average number of bits needed to transmit quantum states.---}
While Result~\ref{ResultCriticalVisibility} provides the classical complexity for fixed message dimension,  it is also interesting to compute the average number of bits required to simulate the set of states in Eq~\eqref{eq:noisy_quantum_theory} using a CCM. Thus, consider that for each shared variable $\lambda$, Alice may encode her state into a classical message using $r_\lambda$ different symbols. The number of transmitted bits on average is therefore
\begin{equation}
\mathbb{E}[\log(r_\lambda)] = \int d\lambda\, q(\lambda) \log(r_\lambda).
\end{equation}
We consider this classical protocol to be optimal, since one can always enlarge the number of symbols adding no further information. It follows from Property 7 in section~\ref{section_ElementaryProperties} that for the set of all pure quantum states, also this average bit-cost is infinite. Using  Result~\ref{ResultCriticalVisibility}, we can go further by largely determining how it scales as a function of the visibility and the dimension.
\begin{corollary}[Average classical complexity]\label{CorollaryAverageBits}
Consider the set of all pure $d$-dimensional states subject to isotropic noise, as given in Eq~\eqref{eq:noisy_quantum_theory}. The average number of bits of communication in a classical simulation is bounded by
\begin{equation}\label{eq:lower_bound_bits}
 (d-1) \log(\frac{1}{1-v}) \leq \mathbb{E}[\log(r_\lambda)] \leq 1+  (d-1) \log(\frac{1}{1-v}).
\end{equation}
\end{corollary}	
\begin{proof}
Denote by $v = \int d\lambda q(\lambda) v_\lambda$ the average visibility at which the states in Eq~\eqref{eq:noisy_quantum_theory} are simulable when using $\log(r_\lambda)$ bits on average.  Here, $v_\lambda=(d\mathcal{I}_{d,r_\lambda}-1)/(d-1)$ and $\mathcal{I}_{d,r_\lambda}$ takes the form in Eq~\eqref{eq:integral_main}. The number of symbols required to achieve $v_\lambda$ is lower bounded by $r_\lambda \geq (1-v_\lambda)^{-(d-1)}$, see Eq~\eqref{eq:upper_bound}. This implies that $\mathbb{E}[\log(r_\lambda)] \geq (d-1) \int d\lambda q(\lambda) (-\log(1-v_\lambda))$. Using Jensen's inequality, we obtain $\mathbb{E}[\log(r_\lambda)] \geq -(d-1) \log(1-v)$.  To get an upper bound,  let $r$ be the number of symbols required to obtain the visibility $v$. From Eq~\eqref{eq:lower_bound} we have that the visibility is lower bounded by $v \geq 1-  \left( \prod_{c=2}^r \frac{c}{1+c}\right)^a \geq 1-  \left( \frac{2}{r+1}\right)^a$, where we have used that $(1+\frac{1}{c})^a \leq 1 + \frac{a}{c}$ with $a = 1/(d-1)$. Hence, it holds that $r < 2 (1-v)^{-(d-1)}$. Consequently, $\mathbb{E}[\log(r_\lambda)] \le \log[r] \leq 1 + (d-1)\log (\frac{1}{1-v})$.  This yields  Eq~\eqref{eq:lower_bound_bits}.
\end{proof}

We see that the average bit-cost of noisy quantum theory  diverges logarithmically in the visibility and linearly in the dimension. By Result~\ref{Result2}, it means the same divergence applies for the generation of quantum correlations in the PAB scenario. This provides an interesting contrast between the PAB scenario and the PM scenario, because while the classical dimension needed to simulate quantum correlations is infinite when $d\geq 4$ \cite{Gois2026}, it is finite when the metric instead is the  average bit-cost \cite{Castellano_2026}. Corollary~\ref{CorollaryAverageBits} therefore shows that this discrepancy is not fundamental to quantum correlations, but rather a consequence of limiting them to the PM scenario. By considering more general experiments, such as the PAB scenario, also the average bit-cost becomes infinite.

\section{Witnessing the complexity of classical models}\label{secWitness}
The previous section showed us the fundamental cost of simulating noisy quantum theory. However, no real-life quantum device generates all states in the theory. Most quantum information protocols are based on preparing a reasonably small number of well-selected states. Here, we show how one can witness the classical complexity of such quantum preparation devices through experimental tests that work in any dimension $d$. While we note that this could in principle be addressed using the SDP criterion in Observation~\ref{Result1}, computing it becomes unpractical  already for quite small $d$. Our approach here is analytical and tailored for devices that aim to prepare either mutually unbiased bases (MUBs) \cite{Wootters_1989, durt_2010} or equiangular tight frames (ETFs) \cite{Lemmens_1973, Renes_2004}. Such devices are usual targets in quantum information; they have previously been considered for witness tests of for example quantum key distribution \cite{Cerf2002}, entanglement detection \cite{Spengler2012, Morelli2023}, measurement incompatibility \cite{Designolle2019} and prepare-and-measure correlations \cite{Brunner2013, Tavakoli2019}.

Let  $\{\ket{e_{a}^z}\}_{a,z}$ be a set of $d$-dimensional bases where $z=1,\ldots,m$ is the basis index and $a=1,\ldots,d$ is the element index. They are called MUBs if $|\langle e_{a}^{z}| e_{a'}^{z'}\rangle|^2 = \frac{1}{d}$ for all $a, a'$ and $z \neq z'$ \cite{Wootters_1989}. Similarly, let $\{\ket{f_a}\}_a$ be a set of states with $a=1,\ldots,n$ with $d\leq n\leq d^2$. They are called an ETF if $\left|\braket{f_a}{f_{a'}}\right|^2 = \frac{n-d}{d(n-1)}$ for all $a\neq a'$, where the right-hand-side is the smallest possible value \cite{welch_1974}. In a nutshell, the MUB property ensures that any state from one basis has a uniform probability over any other basis in the set, while the ETF property means that all pairs of states in the set have the same relative angle and that this angle is as large as possible.

When targeting MUB states, it is intuitive to measure the initially unknown states $\{\rho_{a,z}\}_{a,z}$ in the $z$'th MUB and check the probability that the outcome indeed is $a$. Our witness is the sum-total probability of this expected outcome, namely
\begin{equation}\label{eq:mub_witness}
\mathcal{W}_\mathrm{MUB}=\sum_{z=1}^m\sum_{a=1}^d \langle e_{a}^z| \rho_{a,z} | e_{a}^z \rangle.
\end{equation}
If we truly have MUB states, then we reach the algebraic maximum of $\mathcal{W}_\mathrm{MUB} = m d$.

Similarly, when targeting ETFs, we can project each of the initially unknown states $\{\sigma_a\}_a$ onto the corresponding target element $\ket{f_a}$ in the ETF and consider the sum-total probability of success. This witness becomes
\begin{equation}\label{eq:eam_witness}
\mathcal{W}_\mathrm{ETF} = \sum_{a=1}^n \langle f_{a}| \sigma_{a} | f_{a} \rangle.
\end{equation}
If the states indeed are the ETF, we obtain the algebraic maximum of $\mathcal{W}_\mathrm{ETF}=n$.

Our purpose now is to place upper bounds on the MUB- and ETF-witnesses that are satisfied by any set of quantum preparations whose classical complexity is no larger than $r$. The violation of such a bound would certify that the quantum preparations have classical complexity no smaller than $r+1$. Such bounds are provided by the next result.

\begin{observation}\label{result:witness_mub}
If the $d$-dimensional states $\{\rho_{a,z}\}_{a=1,z=1}^{d,m}$ admit an $r$-complex classical communication model, then the MUB-witness in Eq~\eqref{eq:mub_witness} satisfies
\begin{equation}\label{MubWitnessBound}
\mathcal{W}_\mathrm{MUB} \leq m+\sqrt{m(d-1)(r-1)}.
\end{equation} 

Furthermore, if the $d$-dimensional states $\{\sigma_{a}\}_{a=1}^n$ admit an $r$-complex classical communication model, then the ETF-witness in Eq~\eqref{eq:eam_witness} satisfies
\begin{equation}\label{EAMWitnessBound}
\mathcal{W}_\mathrm{ETF}  \leq \frac{n}{d}\left(1 + (d-1)\sqrt{\frac{r-1}{n-1}} \right).
\end{equation} 
\end{observation}

\begin{proof}
We prove the  bound on the MUB-witness and leave the proof of the ETF-witness for Appendix~\ref{App:ETF}, as it uses similar methods.

Assume that the states $\{\rho_{a,z}\}$ are $r$-complex \eqref{rcomplex}. Because the witness is linear, it then reaches its largest value for a deterministic encoding strategy $\gamma$. This means  $\mathcal{W}_\mathrm{MUB}\leq \max_\gamma \sum_{c=1}^r \lambda_{\max}(O_{c})$, where $\lambda_{\max}(O_{c})$ is the largest eigenvalue of the Hermitian operator $O_{c}=\sum_{a,z} D_\gamma(c|a,z)\ketbra{e_a^z}$. We will  upper-bound $\lambda_{\max}(O_{c})$. For that,  use the inequality $(\tr(O_{c})- \lambda_{\max}(O_{c}))^2 \leq (d-1)(\text{tr} ( O_{c}^2)- \lambda_{\max}(O_{c})^2)$ and solve it for $\lambda_{\max}(O_{c})$ to obtain
\begin{align}\nonumber
\lambda_{\max}(O_{c}) \leq \frac{\tr(O_c)}{d}+\frac{1}{d}\sqrt{(d-1)\left(d\tr(O_c^2)-\tr(O_c)^2\right)}
\end{align}
To compute $\tr(O_c)$ and  $\tr(O_c^2)$, we now use that every term in $O_c$ is either a null or a rank-1 projector. Denote the number of non-zero terms in  $O_c$ associated to given indices $(z,c)$ and $c$ by $k_{z,c}=\sum_a D_\gamma(c|a,z)$ and $k_c = \sum_z k_{z,c}$, respectively, where $k_{z,c} \in [0,d]$ and $k_c \in [0,md]$. Then it follows that $\tr(O_{c}) = k_c$ and $\text{tr} (O_c^2)\leq  k_c+\frac{m-1}{md}k_c^2$, where we have used that $\sum_{z} k_{z,c}^2\geq  k_{c}^2/m$. This results in $\lambda_{\max}(O_{c}) \leq (k_c+\sqrt{k_c(d-1)(d-k_c/m)})/d\equiv g(k_c)$. Hence, we have that $\mathcal{W}_\mathrm{MUB}\leq  \max_\gamma \sum_{c=1}^r g(k_c)$. Lastly, since the function $g$ is concave in $k_c$ we apply Jensen's inequality $\frac{1}{r} \sum_c g(k_c) \leq g(\sum_c k_c/r)$ and $\sum_c k_c = md$. This yields Eq~\eqref{MubWitnessBound}.
\end{proof}

An immediate corollary is that the set of MUB states in $m$ bases, when subjected to  isotropic noise of visibility $v$, satisfies the inequality  
\begin{equation}\label{eq:vis_MUBs}
v_\mathrm{MUB} \leq  \sqrt{\frac{r-1}{m(d-1)}}.
\end{equation}	
Since the visibility monotonically decreases in $m$, we can rule out an $r$-complex model using successively lower visibilities by including more bases.

Another example is from the ETF-witness. We may choose a symmetric informationally complete (SIC) set of states, corresponding to the maximal size $n = d^2$. The  isotropic visibility needed to certify a given classical complexity in this test  becomes
\begin{equation}\label{eq:vis_eam}
v_\mathrm{SIC} \leq \frac {1}{d+1}\left(1+d\sqrt{\frac{r-1}{d^2-1}}\right).
\end{equation}

\section{Algorithmic methods for classical complexity}\label{secAlgo}
Qubits are the standard building-block of quantum information, making them an important special case for our analysis of classical complexity. While the SDP criterion in Observation~\ref{Result1} applies in general, the computation quickly becomes inefficient as we increase the number of states and/or the classical complexity. The scalability bottleneck is that the number of possible compression strategies in the CCM grows exponentially with the size $n$ of the ensemble. For instance, at the lowest possible complexity, $r=2$,  we find the SDP  to no longer be practical for sets of $n\gtrsim 17$ qubit states.  In contrast, it is in principle only necessary to consider a number of strategies growing linearly with $n$ to fully describe an arbitrary CCM. To see this, consider that the noisy qubit set $\mathcal E = \{v \ \vec n_x\}_{x=1}^n$ is $r$-complex for visibilities $v$ below some critical threshold $v^*(\mathcal E)$, and $\vec{n}_x$ is the Bloch vector associated with $x$th state. When $v \leq v^*(\mathcal E)$, such that $\mathcal E\in \mathcal{S}_r$, Carathéodory's theorem implies that $\mathcal E$ can be written as a convex combination of at most $3n+ 1$ extremal elements $\lambda$ in $\mathcal S_r$. That is, there exists a probability distribution $q(\lambda)$ and $r$-complex sets $\mathcal E_\lambda = \{\vec{ \tau}_{x,\lambda}\}_{x}$, with $\vec{ \tau}_{x,\lambda} = \sum_{c=1}^r p(c|x,\lambda) \vec{u}_{c,\lambda}$, such that
\begin{equation}
v \ \vec{n}_x = \sum_{\lambda=1}^m q(\lambda) \sum_{c=1}^r p(c|x,\lambda) \vec{u}_{c,\lambda} \qquad \forall x,
\end{equation}
where $m \leq 3n + 1$.

However, although this tells us that a small cardinality for $\lambda$ is sufficient, the challenge is that we do not know how to select the relevant extreme-points of $\mathcal{S}_r$. In this section, we address this by developing  heuristic programs, both for interior-point search in $\mathcal{S}_r$ and for bounding $\mathcal{S}_r$ from the outside. Through case studies, we show  that these bounds often match up to a small gap, and that the computations are substantially more scalable in both $n$ and $r$ as compared to the SDP in Observation~\ref{Result1}.

\subsection{Inner approximation}
We first put forward a heuristic convex search program for qubit systems that serves as an interior-point method in $\mathcal{S}_r$.  Start by building a large pool, $\mathcal A$, composed of sets of $r$ Bloch vectors $\mathcal U_\lambda = \{\vec u_{1,\lambda},\dots,\vec u_{r,\lambda}\}$ indexed by $\lambda$. The size of the pool and the choice of candidates $\lambda$ determine the quality of the final result. In App~\ref{App:enhancing_SDP} we discuss methods for how to generate a good choice of $\mathcal A$. From the pool $\mathcal A$, we then select a few dozen candidates $\lambda$ and initialise an active set $\mathcal A_0$ \cite{active_set}. We solve the linear program over  $\mathcal A_0$,
\begin{equation}\label{SDPrcomplex_activeset}
	\begin{aligned}
		\max_{\substack{v,\{q(\lambda)\}, \{p(c, \lambda|x)\}}} \quad
		& v\\ 
		\text{s.t.}\quad
		& v \ \vec{n}_x =\sum_{\lambda \in \mathcal{A}_0}\sum_{c=1}^r p(c, \lambda|x) \vec u_{c,\lambda},
		\quad \forall x, \\
		& \sum_{c=1}^r p(c, \lambda|x) = q(\lambda),
		\quad \forall x, \lambda \in \mathcal A_0, \\
		& \sum_{\lambda \in \mathcal A_0} q(\lambda) = 1 \\
		& p(c, \lambda|x) \geq 0, 
		\quad \forall c,  x, \lambda\in \mathcal A_0
	\end{aligned}
\end{equation}
This puts a cheap, very loose but non-trivial lower bound on the critical visibility\footnote{For a good initial set $\mathcal{A}_0$, the LP returns a non-zero visibility. However, if $v \leq 0$, we add roughly ten more candidates $\lambda$ to $\mathcal A_0$ and retry.}. Next, we write the dual of \eqref{SDPrcomplex_activeset}, namely
\begin{equation}\label{SDPrcomplex_activeset_dual}
	\begin{aligned}
		\min_{t,\{\nu_{\lambda,x}\}, \{\vec{w}_x\}} \quad
		& t\\ 
		\text{s.t.}\quad
		& 1 =  \sum_{x=1}^n\langle \vec w_x,\vec n_x \rangle, \\
		& \langle \vec w_x,\vec u_{c,\lambda} \rangle \leq -\nu_{\lambda,x}, \quad  \forall x, c, \lambda \in \mathcal A_0, \\
		& \sum_{x=1}^n \nu_{\lambda,x} + t = 0, \quad \forall \lambda \in \mathcal A_0.
	\end{aligned}
\end{equation}
When $\mathcal E \notin \mathcal S_r$ the dual LP provides a procedure to check if we caught all the useful choices of $\lambda$ in $\mathcal A$. Denote by $\beta_\lambda \equiv \sum_x \max_c \langle \vec w_x, \vec u_{c,\lambda} \rangle$. By inserting the definition of $r$-complexity in the primal LP to the dual constraint $1 = \sum_x  \langle \vec w_x,\vec n_x \rangle$, we obtain that the optimal value of the primal LP is upper bounded by $v \leq \sum_{\lambda} q(\lambda) \beta_\lambda$ for all $r$-complex CCMs using $\lambda \in \mathcal A_0$. By constructing the dual variables $\vec w_x$ from the solution of our primal LP and evaluating $\beta_\lambda$ for every $\lambda$ in $\mathcal A$ but not in $\mathcal A_0$, we then add the candidates $\lambda$ that give the largest value $\beta_\lambda > v$ to our active set, updating it from $\mathcal{A}_0$ to $\mathcal{A}_1$. Typically, we add around a few dozen candidates $\lambda$ per iteration. Thereafter, we compute the LP \eqref{SDPrcomplex_activeset} over $\mathcal A_1$ and then add new candidates to the active set by checking $\beta_\lambda$, which is updated to $\mathcal A_2$ etc. We also prune candidates $\lambda$ that have been found to have a zero-weight $q(\lambda)$ by the LP over a few iterations. This keeps the size of the active set small. We then repeat this procedure until either  $\beta_\lambda = v$ for all $\lambda$ or the algorithm  converges to a local optimum. Note that each LP computes a lower bound on the critical visibility, i.e., $v \leq v^*$. While we focus on qubit states, this method can be straightforwardly generalised to any dimension.


\subsection{Outer approximation}
We develop a method to compute outer bounds on $\mathcal{S}_r$ for  qubit sets $\mathcal E$. It is based on two steps; (i) a heuristic method to find a good witness candidate for ruling out complexity $r$, and (ii) turning the candidate witness into a certificate.

\subsubsection{Constructing the witness}
We start with a heuristic method to find a witness candidate suitable for showing separation between  $\mathcal E$ and $\mathcal{S}_r$. To this end, consider the equality constraint  $\sum_{x=1}^n \langle \vec w_x, \vec n_x \rangle = 1$ from the dual LP. The largest value that this expression can achieve for any $r$-complex CCM is given by
\begin{equation}\label{eq:witness}
h(w) = \max_{\vec u_1,\dots,\vec u_r} \sum_x \max_c \, \langle \vec w_x, \vec u_c\rangle.
\end{equation}
From LP duality it follows that any such witness $h(w)$ is an upper bound on the critical visibility $v^*(\mathcal E,r) \leq  h(w)$, where equality is attained by solving the dual LP over all $r^n$ possible strategies $\lambda$. Since it is computationally costly to exactly solve the  constrained optimisation problem, we instead relax the problem and use an alternating heuristic search  to find a good witness candidate. 

The alternating routine is decomposed into two sub-routines. First, for fixed $w$, the program searches for a good set of $r$ Bloch vectors $u = (\vec u_1,\dots, \vec u_r)$ that yields a large witness $h(w)$. This is done by using a type of  clustering method \cite{clustering}, where the main idea is to split the Bloch vectors in $w$ into $r$ groups, such that all vectors in the same group are close in distance. Each group is then assigned a point $\vec u_c$, which is taken to be the normalised sum of the group element. The groups are iteratively updated until they reach a steady state. In the second sub-routine, we keep the final $u$ from the clustering method fixed. To minimise $h(w)$ with respect to $w$, we update $w$ along the negative direction of its subgradient $g = \partial h(w)$ that is orthogonal to $\{\vec n_x\}_x$, such that the updated witness remains properly normalised. The sub-routines are iterated until a local optimum is reached. See Appendix~\ref{App:enhancing_SDP} for details on how the program was implemented.

However, the witness $h(w)$ value obtained via the heuristic search algorithm is not a certified upper bound on the critical visibility. The reason is that $h(w)$ is not optimised over all possible sets of $u$ as stipulated in Eq~\eqref{eq:witness}. Nevertheless,  we show next that it is still possible to obtain a certificate from the witness $w$. We provide two types of certificate, denoted by $A$ and $B$, where the latter is typically best when $r\sim n$ and the former is typically best otherwise. 

\subsubsection{Certificate A: polytope approximation and Lagrangian bound}
Certificate $A$ uses a polytope approximation of the Bloch sphere and a Lagrangian relaxation method. To this end, choose some finite set of $N$ Bloch vectors $Y = \{\vec y_j\}_{j}$ of unit length. The points on the Bloch sphere's surface can be interpreted as the vertices of a polyhedron embedded in the Bloch sphere. How well the polyhedron approximates the Bloch sphere is quantified by the parameter $s \in (0,1)$, which is defined as the radius of the largest sphere that lies entirely within the convex hull of $Y$ \cite{AB_Banach}. The largest value the witness $w$ in Eq~\eqref{eq:witness} can reach over the discrete set $Y$ is given by 
\begin{equation}\label{eq:witness_discrete}
h_Y(w) = \max_{T \subset Y \text{ and } |T|\leq r} \sum_x \max_{j \in T} \langle \vec w_x, \vec y_j\rangle,
\end{equation}
where $T = \{\vec y_j\}_j$ has size at most $r$. While $h(w) \geq h_Y(w)$,  it holds that $h(w) \leq \frac{1}{s}h_Y(w)$ \cite{Liu_2025}. The latter can be seen by noting that elongating each Bloch vector $\vec y_j \in Y$ by a factor $s^{-1}$ results in that the convex hull of $\frac{1}{s}Y$ contains the Bloch sphere. However, solving the optimisation problem in Eq~\eqref{eq:witness_discrete} quickly becomes inefficient, since it requires selecting an optimal set $T$ of $r$ vectors $\{\vec y_j\}_{j \in T}$ among $\binom{N}{r}$ different combinations. Instead, we relax the problem and compute an upper bound on $h_Y(w)$. 
\begin{observation}[Lagrangian certificate]\label{result:L_certificate}
The optimal value of $h_Y(w)$ in Eq~\eqref{eq:witness_discrete} given the polytope approximation $Y$ is upper bounded by
\begin{equation}
\hat h_Y(w) =\min_{\mu_1,\dots,\mu_n} \left( \sum_{x=1}^n  \mu_x + \sum_{j = 1}^r G_j(\mu) \right),
\end{equation}
where $\mu_x \geq 0$ and $G_j(\mu) = \sum_{x=1}^n \max[\langle \vec w_x,\vec y_j\rangle -\mu_x,0]$ is sorted in descending order such that $G_1(\mu) \geq G_2(\mu) \geq \dots $.
\end{observation}
\begin{proof}
Let $g_x(j) = \langle \vec w_x, \vec y_j \rangle$ and introduce the Lagrangian multipliers $\mu_x \geq 0$. Then it holds that 
$\max_{j\in T} g_x(j)  \leq \mu_x + \sum_{j\in T}\max(g_x(j)-\mu_x,0)$. Sum both sides of the inequality over the settings $x$ and sort $G_1(\mu) \geq G_2(\mu) \geq \dots $. Optimising the expressions over the subset $T \subset Y$ of cardinality $r$ yields $h_Y(w) \leq  \sum_{x=1}^n  \mu_x + \sum_{j = 1}^r G_j(\mu)$. Since the upper bound is valid for all sets of non-negative $\mu_x$, the tightest bound is obtained by minimising over all such choices.
\end{proof}

This certification gets more expensive as we increase $r$, because more terms $G_j(\mu)$ are needed. In App~\ref{App:enhancing_SDP} we describe how to numerically compute $\hat h_Y(w)$. Broadly, the implementation uses the Bloch sphere approximation introduced in Ref~\cite{Liu_2025} and performs the minimisation over $\mu$ using a subgradient method \cite{Subgradient_2014}.

\subsubsection{Certificate B: merge bound}
For certificate $B$, consider again the witness $h(w)$ in Eq~\eqref{eq:witness}. Note that each of the $n$ witness vectors $\vec w_x$ picks the candidate $\vec u_c$ which it overlaps with the most. This splits the $n$ elements in $w$ into $r$ disjoint groups $G_c$,  each associated with a distinct vector $\vec u_c$. In fact, optimising $h(w)$ over the set $u$ is equivalent to finding the optimal grouping $G_c$ that maximises the sum of the lengths of the group sum, i.e.,
\begin{equation}\label{eq:h_group}
	h(w) = \max_{G_1 \sqcup \dots \sqcup G_r} \sum_{c=1}^r \big\|\! \sum_{x \in G_c} \vec w_x \big\|.
\end{equation}

If each vector $\vec w_x$ had its own group $G_x$, we could trivially reach the algebraic maximum value $ \sum_x \|\vec w_x\|$. However, the partition of $n$ distinct elements into $r$ groups naturally results in that this value cannot be reached. To quantify how much we lose by merging a set of elements $\vec w_x$ into the same group $G$ we introduce the loss-term
\begin{equation}
	\text{Loss}(G) =  \sum_{x\in G} \|\vec w_x\| -\|\sum_{x\in G} \vec w_x\|.
\end{equation}
Then, $h(w)$ can be expressed as follows
\begin{equation}\label{eq:h_loss}
	h(w) = \sum_{x=1}^n \|\vec w_x \| - \min_{G_1\sqcup \dots \sqcup G_r} \sum_{c=1}^r \text{Loss}(G_c),
\end{equation}
where the minimisation is performed over all possible disjoint groupings. By bounding the total loss term from below, we can obtain an upper bound on the witness.

\begin{observation}[Merge certificate]\label{result:merge_certificate}
The value of any function $h(w)$ of the form in Eq~\eqref{eq:h_loss} is upper bounded by
\begin{equation}\label{eq:upper_merge}
h(w) \leq \sum_{x=1}^n \|\vec w_x \| - \sum_{k = 1}^K L_{i < j}^{(k)},
\end{equation}
where $L_{i < j} = \|\vec w_i\|+\|\vec w_j\| -\| \vec w_i+\vec w_j\|$ is the pairwise loss term associated with merging elements $i$ and $j$ into the same group and we choose the $K = \lceil\frac{n-r}{2} \rceil$ smallest pairs, i.e., $L^{(1)} \leq L^{(2)} \leq \dots$.
\end{observation}
\begin{proof}
Consider a group $G$ that contains two disjoint groups $A$ and $B$. Then it follows that $\text{Loss}(A \sqcup B) \geq \text{Loss}(A) + \text{Loss}(B)$, meaning that the loss is super-additive. Hence, for a set of $r$ groups $\{G_c\}$, where each group contains $m_c$ elements, there are at least $K = \lceil \frac{n-r}{2} \rceil$ such disjoint pairs across all groups. Thus, compute the pairs $L_{i < j} = \|\vec w_i\|+\|\vec w_j\| -\| \vec w_i+ \vec w_j\|$ for $i,j = 1,\dots,n$ and sort them in increasing order $L^{(1)} \leq L^{(2)} \leq \dots$. By selecting the $K$ smallest pairs we obtain the lower bound 
\begin{equation}
\min_{G_1\sqcup \dots \sqcup G_r} \sum_{c=1}^r  \text{Loss}(G_c) \geq \sum_{k = 1}^K L_{i < j}^{(k)}.
\end{equation}
\end{proof}

In general, the upper bound in Eq~\eqref{eq:upper_merge} is loose because the $ K$ smallest pairs $L_{i <j}$ might not be mutually disjoint. This means that this bound typically is good in the regime $r \sim n$, where the probability that an element $\vec w_x$ contributes to two different loss-terms is small.

\begin{table}[t!]
	\centering
	\begin{tabular}{cc|cccc}
		\toprule
		$n$ & $r$ &  $v_{\text{lb}}$  & $v_{\text{ub}}$  & $t_{\text{lb}} (s)$ & $t_{\text{ub}} (s)$\\
		\midrule
		100 & 4   & 0.7459 & 0.7549 & 16  & 5   \\
		100 & 8   & 0.8704 & 0.8811 & 30  & 11  \\
		\addlinespace
		10 000	& 4 & 0.7400 &	0.7581 & 8030 & 104 \\
		10 000	& 8 & 0.8617  &	0.8824 & 17 580 & 180 \\
		\addlinespace
		150 & 75  & 0.9860 & 0.9927 & 356 & 380 \\
		150 & 110 & 0.9912 & 0.9976 & 436 & 21  \\
		150 & 149 & 0.9957 & 0.9999 & 364 & 4   \\
		\addlinespace
		500 & 250 & 0.9956 & 0.9986 & 2640 & 930 \\
		500 & 400 & 0.9978 & 0.9994 & 2870 & 754 \\
		500 & 490 & 0.9996 & <1 & 6290 & 160  \\
		\bottomrule
	\end{tabular}
	\caption{Results for sets of $n$ Bloch vectors on a Fibonacci lattice for different $r$. The time $t$ needed to compute the lower $v_{\text{lb}}$ and upper $v_{\text{ub}}$ bounds on the critical visibility is given in seconds. }	\label{tab:near-uniform-main}
\end{table}

\subsection{Case studies}
We demonstrate the usefulness of the inner and outer approximations through relevant case-studies. Let us consider a set of $n$ Bloch vectors $\mathcal E = \{\vec n_x\}_x$ in Fibonacci lattice \cite{Fibonacci_lattice}, which makes them near-uniform. For different values of $n$ and $r$ we compute lower and upper bounds on the critical visibility for an $r$-complex model, thereby confining the true threshold to the interval between the two bounds, i.e., $v_{\text{lb}} \leq v^*(\mathcal E) \leq v_{\text{ub}}$. As showcased in Table \ref{tab:near-uniform-main}, the numerical methods are both computable in reasonable time, provide a good estimate of the critical visibility and can be evaluated for large sets of states. In particular, for fixed $r$ the ability of CCMs to simulate sets of qubit states remains relatively constant, even as the set size grows by two orders of magnitude. Moreover, already for $r/n \sim 0.75$ the lower and upper bound are above $99\%$. This suggests that the classical cost of simulating quantum transmission truly becomes very expensive only in the limit of nearly-pure qubit states. Moreover, by comparing the time elapsed when computing the inner and outer approximation, we note that the former in general takes longer than the latter. This is because the former needs to compute a large pool of good candidates and it solves an LP in each iteration. In the regime where $r \ll n$, we see that the best upper bound is obtained using certificate A, whereas for $r \sim n$ it is obtained using certificate B.

\section{Conclusions}
The classical cost of simulating the black-box input-output predictions of quantum communication experiments  has received much research attention  (see e.g.~\cite{Buhrman2010, Brask2026, Oliveira2026}), but the classical cost of simulating the quantum communication itself  has remained largely unaddressed. Here, we have introduced a theoretical framework for filling this gap, centered about the notion of \textit{classical complexity}, which quantifies the minimum  amount of classical communication needed to reproduce quantum state transmission. It provides a task-independent notion of quantum communication advantage, addressed directly at the quantum states transmitted in the experiment.

A conceptually central result is that, while designed for accounting for the classical cost of simulating quantum states, the classical complexity also gives a complete characterisation of whether quantum states can be used for producing quantum correlations in appropriate black-box input-output experiments (Result~\ref{Result2}).  This one-to-one connection between quantumness on the level of the states and quantumness on the level of device-independent predictions provides our models with further  operational significance. One may view this as the communication-experiment counterpart to the role that entanglement plays in Bell-type experiments \cite{Brunner2014, Tavakoli2022}.

This connection provided additional motivation for studying the classical cost of simulating noisy quantum theory, which we have largely resolved in Result~\ref{ResultCriticalVisibility}. Our main conclusions and observations following this cost characterisation  are the following. 
\begin{enumerate}
	\item Recent works have studied the cost of simulating quantum correlations in prepare-and-measure scenarios \cite{Renner2023, Gois2026}  (see also \cite{Castellano_2026, Tavakoli2026}) and found that whether the cost is finite or infinite ultimately depends critically on the dimensionality. However, our result shows a simpler relation: the cost is infinite already for qubits. The reason this was not revealed in the previous works is their restriction to the prepare-and-measure scenario, which is the simplest possible experiment in which quantum advantage from communication can come about. Our result therefore shows that the full power of the quantum advantage inherent to a given set of quantum states is revealed only by studying the more sophisticated experiments, such as the prepare-and-broadcast scenario.

	\item Both dimensionality and purity have a crucial role in making it expensive to classically simulate the quantum advantage. Increasing the dimensionality makes classical models exponentially more expensive, and reducing the noise increases the cost as a $(d-1)$-degree power-law. Dimensionality can therefore be viewed as a powerful resource for quantum advantage, both in state transmission and, due to Result~\ref{Result2}, also in quantum correlations.
	
	\item Even though this classical simulation cost explicitly accounts for noise, it is only an in-principle (fundamental limit) result when it comes to considering experimental tests aiming to demonstrate a very high classical cost on the level of black-box input-output correlations. That is because Result~\ref{Result2} exploits auxiliary self-testing in order to convert the quantum advantage in state transmission into quantum advantage in correlations. It means that in order to experimentally access even a part of the fundamentally very large advantage revealed here, one must consider much more practically designed protocols. It presents a future direction of research.
\end{enumerate}

In addition, we have developed efficient tools for certifying classical complexity in more practical quantum devices, which are usually aimed at preparing certain types of particularly relevant states. In section~\ref{secWitness} we did this based on witness tests tailored for high-dimensional mutually unbiased bases and equiangular states, both of which are commonplace targets for quantum devices. In section~\ref{secAlgo}, we focused on qubits and introduced algorithms that can analyse many hundreds and even several thousands of qubit states and put reasonably tight bounds on their classical complexity. Also from these computational tools, we see the same trend observed for the fundamental cost of simulating full quantum theory, namely that classical models can somewhat  approximate quantum states using only low complexity, but that this cost  starts to  rapidly increase when we require the approximation to be increasingly accurate. This suggests that very-low-noise quantum communication devices may be very much harder to classically simulate than devices that operate at good but more commonplace noise levels. It presents an interesting future opportunity for large-scale quantum advantages in communication.

\begin{acknowledgements}
We thank Gabriele Cobucci for discussions. A.T.~thanks Rafael Chaves for his hospitality and stimulating discussions in Natal, and Carles Roch i Carceller for numerous  discussions that helped to shape and inspire this work. The project is financially supported by the Swedish Research Council under Contract No. 2023-03498, the  Swedish Foundation for Strategic Research, by the Knut and Alice Wallenberg Foundation through the Wallenberg Center for Quantum Technology (WACQT),  by the Crafoord Foundation, the Krapperup Foundation and by the Royal Physiographic Society of Lund.

\textit{AI statement.---} We used ChatGPT 5.6 Sol to polish parts of our text and to obtain the upper and lower bounds on the integral in \eqref{eq:integral_main} appearing in Result~\ref{ResultCriticalVisibility}. The correctness of these bounds was verified by us and their presentation was adapted for accessibility to human readers. We further used Claude Opus 4.5 to advise in the construction of the algorithms in section~\ref{secAlgo}. The implementation of all relevant codes was the work of the authors. 
\end{acknowledgements}

\bibliography{references}

\appendix
\onecolumngrid

\section{Witness tests for $r$-complex models}\label{AppPrimalDual}
In this section we show how to derive a witness test for $r$-complex models whenever $r\ge 2$. To this end, we start by recasting the feasibility SDP in Eq~\eqref{SDPrcomplex} as a strictly feasible form as follows
\begin{equation}\label{SDPrcomplex_strict}
	\begin{aligned}
		\max_{t,\{\tilde{\sigma}_{c,\mu}\}_{c,\mu}}\quad
		& t \\ 
		\text{s.t.}\quad
		& \rho_x
		= \sum_{\mu=1}^{r^n}\sum_{c=1}^{r}
		D_\mu(c|x)\tilde{\sigma}_{c,\mu},
		&& \forall x, \\
		& \tr(\tilde{\sigma}_{c,\mu})=\tr(\tilde{\sigma}_{1,\mu}),
		&& \forall c,\mu, \\
		& \tilde{\sigma}_{c,\mu}\succeq t\mathds{1},
		&& \forall c,\mu .
	\end{aligned}
\end{equation}
To obtain the strictly feasible SDP formulation we have relaxed the problem and allow for the set of variables $\{\tilde \sigma_{c,\mu}\}_{c,\mu}$ to have negative eigenvalues. This is done by replacing the constraint $\tilde{\sigma}_{c,\mu} \succeq 0$ by $\tilde{\sigma}_{c,\mu} \succeq t \openone$, where $t$ is the new variable. By maximising over $t$, we then search for the set of Hermitian states, consistent with the data, with the largest eigenvalues. If and only if $t^*\geq 0$, where $t^*$ is the solution to the SDP, the original SDP is feasible and $\mathcal{E}\in\mathcal{S}_r$ \cite{Skrzypczyk_2023}. 

From the strictly feasible form of the primal SDP we can derive its dual formulation using the associated Lagrangian of the optimisation problem. Specifically, the Lagrangian is obtained by adding to the objective function the constraints multiplied with the dual variables. Here, the Lagrangian function for the SDP reads
\begin{equation}\label{eq:Lagrangian_rcomplexity}
	\begin{aligned}
		\mathcal L &= t + \sum_{x=1}^n \tr(Y_x\left[\rho_x - \sum _{\mu=1}^{r^n} \sum_{c=1}^r D_{\mu}(c|x) \tilde \sigma_{c,\mu}\right]) + \sum _{\mu=1}^{r^n} \sum_{c=1}^r \nu_{c,\mu}\left(\tr(\tilde \sigma_{c,\mu}) -\tr(\tilde \sigma_{1,\mu}) \right) +  \sum _{\mu=1}^{r^n} \sum_{c=1}^r \tr(Z_{c,\mu} \left[\tilde \sigma_{c,\mu}-t\openone\right]) \\
		& = t\left(1-  \sum _{\mu=1}^{r^n} \sum_{c=1}^r \tr(Z_{c,\mu})\right) +  \sum _{\mu=1}^{r^n} \sum_{c=1}^r \tr(\tilde \sigma_{c,\mu}\left[\nu_{c,\mu} \openone + Z_{c,\mu}-\sum_x D_\mu(c|x) Y_x \right]) -  \sum _{\mu=1}^{r^n} \tr(\tilde \sigma_{1,\mu}) \sum_{c=1}^r \nu_{c,\mu} + \sum_{x=1}^n \tr(Y_x \rho_x)
	\end{aligned}
\end{equation}
Here, we have introduced the Hermitian dual variables $Y_x$, and $Z_{c,\mu}$, as well as the scalar variables $ \nu_{c,\mu}$. Specifically, we impose that $Z_{c,\mu} \succeq 0$ to ensure that the Lagrangian is never smaller than the objective function for any feasible solution $t$, i.e., $\mathcal L \geq t$ for all $Z_{c,\mu} \succeq 0$ \cite{Skrzypczyk_2023}. From the Lagrangian we can then extract the dual optimization program 
\begin{equation}
	\begin{aligned}
		\min_{\{Y_x\}_x, \{Z_{c,\mu}\}_{c,\mu}, \{\nu_{c,\mu}\}_{c,\mu}}\quad
		& \sum_x \tr(Y_x\rho_x) \\ 
		\text{s.t.}\quad
		& \sum_x D_\mu(c|x)Y_x
		= \nu_{c,\mu}\mathds{1} + Z_{c,\mu},
		&& \forall c,\mu, \\
		& \sum_{c=1}^{r}\nu_{c,\mu}=0,
		&& \forall \mu, \\
		& \sum_{\mu=1}^{r^n} \sum_{c=1}^r \tr(Z_{c,\mu}) = 1, \\
		& Z_{c,\mu} \succeq 0,
		&& \forall c,\mu. \\
	\end{aligned}
\end{equation}
We notice that the objective function does not depend upon the slack variable $Z_{c,\mu}$. The final simplified SDP is therefore
\begin{equation}
	\begin{aligned}
		\min_{\{Y_x\}_x,\{\nu_{c,\mu}\}_{c,\mu}}\quad
		& \sum_x \tr(Y_x\rho_x) \\ 
		\text{s.t.}\quad
		& \sum_x D_\mu(c|x)Y_x
		\succeq \nu_{c,\mu}\mathds{1},
		&& \forall c,\mu, \\
		& \sum_{c=1}^{r}\nu_{c,\mu}=0,
		&& \forall \mu, \\
		& r^n\sum_x\tr(Y_x)=1, & &
	\end{aligned}
\end{equation}
The first constraint ensures that all $r$-complex communication models return nonnegative values \cite{Tavakoli_2024}.  By considering the dual formulation of the SDP \eqref{SDPrcomplex} we can obtain witness criteria for falsifying $r$-complex models. This can be seen by inserting the definition for $r$-complexity into the objective function of the dual and using the associated constraints, i.e.,
\begin{align}\nonumber
	\sum_x \tr(Y_x\rho_x)& = \sum_x \tr(Y_x \sum_{\mu,c} D_\mu(c|x)\tilde{\sigma}_{c,\mu})	= \sum_{\mu,c} \tr(\tilde{\sigma}_{c,\mu}\sum_x D_\mu(c|x) Y_x )\\
	& \geq \sum_{\mu,c}\nu_{c,\mu} \tr(\tilde{\sigma}_{c,\mu} ) = \sum_{\mu,c}\nu_{c,\mu} \tr(\tilde{\sigma}_{1,\mu} ) = \sum_\mu \tr(\tilde{\sigma}_{1,\mu} ) \sum_c \nu_{c,\mu} =0.
\end{align}
In other words, we can apply the definition of the $r$-complex model to the objective function and use the dual constraints to obtain the inequality
\begin{equation}
	\sum_x \tr(Y_x\rho_x) \geq 0,
\end{equation}
which holds for all $r$-complex models. Thus, the dual variables $\{Y_x\}_x$ provide the observables needed to rule out an $r$-complex model.

\section{Reduction of deterministic strategies in Observation~\ref{Result1}}\label{AppSterlingStrategies}
The number of deterministic strategies appearing in the SDP in Eq~\eqref{SDPrcomplex} scales exponentially in $n$ and polynomially in $r$, namely as $r^n$. This can be reduced w.l.g. To this end, consider two different encoding functions for Alice, $c=f(x)$ and $c'=g(x)$.  Consider now that there exists a bijective function $h$ such that $g=h\circ f$.  If this is the case, whenever the parties have agreed to apply $f$, Bob can compute the permuted version $h(c)$ in his lab and obtain also the message Alice would have sent if they had chosen to encode with $g$ (namely $c'$). Thus, there is no need to separately account for the encoding strategy $g$.

To formalise this idea, we must count the appropriate equivalence classes of the encoding functions. To this end, we first notice that $f(x)=f(x')$ if and only if $g(x)=g(x')$, so we must group in the same class all functions that fulfil this property.  Counting all equivalence classes then reduces to counting the ways the input set is partitioned into subsets. Indeed,  for any message $c$ the set of inputs encoded in that message through a function $f$ is $f^{-1}(c)=\{x: f(x)=c\}$,  which is disjoint from any other $f^{-1}(c')$ for $c'\neq c$.  Evaluating $f^{-1}$ for all $c=1,\dots,r$ gives a partition of the input set into $r$ sets.  Note that any other function $g$ such that $g=h\circ f$ with $h$ a bijection satisfies $g^{-1}(c')=\{x: g(x)=c'\}=\{x: f(x)=h^{-1}(c')=c\}=f^{-1}(c)$ and therefore produces exactly the same partition.  Let us denote the number of non-empty sets in the partition by $k$.  Hence,  in order to get the total number of strategies, we must count the number of ways the $n$ input elements can be partitioned into $k$ non-empty subsets and then sum for all possible values $k=1,\dots,r$. The first quantity corresponds to the Stirling number of the second kind, denoted $S(n,k)$, while the total sum
\begin{equation}\label{eq:stirling}
	B_r(n) \equiv \sum_{k=1}^{r} S(n,k),
\end{equation}
is known as the $r$-partial Bell number. It is the number of deterministic strategies we need to consider in the SDP.

For a fixed $r$ and large $n$, the ratio between the reduced number of strategies and the original number of strategies scales as
\begin{equation}
	\lim_{n\rightarrow \infty}	\frac{B_r(n)}{r^n}= \frac{1}{r!}.
\end{equation}
Thus, we can expect to gain a factor of roughly $r!$ in the number of deterministic strategies. This becomes significant for large values of $r$, which are typically those relevant to simulating quantum states.

Notably, this argument applies directly also to the standard dimension-restricted prepare-and-measure scenario, where Alice and Bob are treated as uncharacterised devices and one is interested in determining the membership of their input-output correlations to the polytope of classical correlations \cite{Brask2026, Oliveira_2026}. Methods of speeding up the membership problem in that problem have been developed \cite{Renner2023}, but to the best of our knowledge the above reduction has not been considered in the topic-specific literature. The above reduction therefore offers a potential further speed-up when appropriately combined with the state-of-the-art.

\section{Device-independent certification proof}\label{AppDI}
We first introduce the self-test proposed in Refs~\cite{Bowles_2018_prl,Bowles_2018_pra} and then use it to show that there exists a black-box type of inequality in the prepare-and-broadcast scenario that is satisfied by all $r$-complex preparations but can be violated by those that do not admit a CCM.

\subsection{Bowles' self-test for qubits}
The self-testing protocol proposed in Refs~\cite{Bowles_2018_prl,Bowles_2018_pra} is based on the Bell scenario in which two parties, Bob and Charlie, aim to maximally violate the triple CHSH inequality
\begin{equation}\label{eq:Bell_triple}
	\begin{aligned}
		\mathcal B &= E_{1,1} + E_{2,1} + E_{1,2} - E_{2,2} \\
		&\qquad+ E_{3,1} + E_{4,1} - E_{3,3} + E_{4,3} \\
		&\qquad \qquad+ E_{5,2}+ E_{6,2} - E_{5,3} + E_{6,3} \leq 6.
	\end{aligned}
\end{equation}
Here, the correlators read $E_{y,z} = \sum_{\tilde b, \tilde c = \pm 1} \tilde b \tilde c\, p(\tilde b, \tilde c|y,z)$. Each row of the inequality corresponds to a standard CHSH parameter, which has maximum quantum value $2 \sqrt 2$, and consequently, the maximal violation of the triple CHSH inequality \eqref{eq:Bell_triple} is $\mathcal B = 6\sqrt{2}$. This value can be achieved when Bob and Charlie share the EPR state $\ket{\phi^+} = \frac{1}{\sqrt{2}}(\ket{00}+\ket{11})$ and Charlie measures the three mutually anti-commuting Pauli observables
\begin{equation}
	\sigma_Z, \sigma_X, \sigma_Y \quad z =1,2,3
\end{equation}
and Bob measures the six rotated Pauli observables 
\begin{equation}
	\frac{\sigma_Z \pm \sigma_X}{\sqrt{2}}, \frac{\sigma_Z \pm \sigma_Y}{\sqrt{2}}, \frac{\sigma_X \pm \sigma_Y}{\sqrt{2}} \quad y =1,2,3,4,5,6.
\end{equation}
This choice of shared state and local measurements is indeed the unique maximiser (up to local isometries and correlated,  auxiliary-controlled transpositions) of the Bell inequality \eqref{eq:Bell_triple}. This is captured by the following self-testing statement.

\begin{lemma}[Bowles' self-testing statement \cite{Bowles_2018_prl}]\label{lemma:bowels_selftest}
	Let Charlie and Bob share the state $\ket{\psi} \in \mathcal H_{C_0} \otimes \mathcal H_{B_0}$ and denote by $Z^{C_0}, X^{C_0}, Y^{C_0}$ three $\pm 1$ outcome observables for Charlie. If one observes a Bell inequality violation of $\mathcal B = 6\sqrt 2$ \eqref{eq:Bell_triple}, then there exist local auxiliary states $\ket{00} \otimes \ket{00} \in [\mathcal H_{C''}\otimes \mathcal H_{C'}] \otimes [\mathcal H_{B''}\otimes \mathcal H_{B'}]$ and a local unitary $U = U_{C} \otimes U_{B}$ where $\mathcal H_B = \mathcal H_{B_0} \otimes \mathcal H_{B'} \otimes \mathcal H_{B''}$, and similarly for $\mathcal H_C$, such that
	\begin{equation}
		\begin{aligned}
			&U[\ket{\psi} \otimes \ket{00} \otimes \ket{00}] = \ket{\xi} \otimes \ket{\phi^+}^{C'B'}\\
			&U[X^{C_0} \ket{\psi} \otimes \ket{00} \otimes \ket{00}] = \ket{\xi} \otimes \sigma_X^{C'} \ket{\phi^+}^{C'B'}\\
			&U[Z^{C_0} \ket{\psi} \otimes \ket{00} \otimes \ket{00}] = \ket{\xi} \otimes \sigma_Z^{C'} \ket{\phi^+}^{C'B'} \\
			&U[Y^{C_0} \ket{\psi} \otimes \ket{00} \otimes \ket{00}] = \sigma_Z^{C''} \ket{\xi} \otimes \sigma_Y^{C'} \ket{\phi^+}^{C'B'}
		\end{aligned}
	\end{equation}
	where $\ket \xi$ takes the form
	\begin{equation}
		\ket{\xi} = \ket{\xi_0}^{C_0 B_0} \otimes \ket{00}^{C''B''} +\ket{\xi_1}^{C_0 B_0} \otimes \ket{11}^{C''B''}.
	\end{equation}
\end{lemma}
Here the superscript denotes the Hilbert space $\mathcal H$ on which an operator acts nontrivially. Since we impose no dimension constraint on the Hilbert spaces $\mathcal H_C$ and $\mathcal H_B$ we can w.l.g.  take the state $\Psi$ to be pure and Charlie's observables to be projective. This follows from the fact that any arbitrary mixed bipartite state can be purified on enlarged local Hilbert spaces and Naimark's dilation of the measurements.

The self-testing statement proves that there exists a local unitaries and auxiliary states such that the target state $\ket{\psi'} \equiv \ket{\phi^+}$ can be extracted from the physical state $\ket{\psi}$ into the $\mathcal H_{C'} \otimes \mathcal H_{B'} $ space. Due to the freedom of local unitary operations one can w.l.g.  express the target state with real numbers \cite{Bowles_2018_pra}. The junk state $\ket \xi$ collects the possible additional degrees of freedom and satisfies the normalisation condition $\braket{\xi_0}+\braket{\xi_1} = 1$.

The self-test further certifies that Charlie's measurement operators $Z^{C_0}, X^{C_0}, Y^{C_0}$ are equivalent to the target observables $(\sigma_Z,\sigma_X,\pm \sigma_Y)$ acting on his local share of $\ket{\phi^+}$. However, since $\sigma_Y$ cannot be expressed using real numbers alone, it is only certified up to possible transposition \cite{Bowles_2018_pra}. This is because quantum correlations are invariant under global transposition, i.e., $\braket*{\psi'| B \otimes C}{\psi'} = \braket*{\psi'| B^T \otimes C^T}{\psi'}$ assuming that the target state $\ket{\psi'}$ is real. To deal with this possibility, the complex observable $\sigma_Y$ has an additional measurement $\sigma_Z$ on the $C''$ space, which functions as a "controlled transposition" measurement. Thus, the measurement $Y^{C_0}$ can be understood as first measuring $\sigma_Z^{C''}$ on $\ket \xi$ and then, depending on the measurement outcome, measuring $\ket{\phi^+}$ on space $C'$ using either the target observable $\sigma_Y$ or its transposition $\sigma_Y^T$ with some unknown probability depending on $\braket{\xi_j}$ with $j= 0,1$. On the support of the joint state $\ket{\psi} \otimes \ket{00} \otimes \ket{00} = U^\dagger (\ket{\xi} \otimes \ket{\phi^+})$, Charlie's projector can be written as \cite{Bowles_2018_pra}
\begin{equation}\label{eq:Pi}
	\Pi_{\tilde c|z}^{C}= U_{C}^\dagger\left(\openone^{C_0}\otimes \sum_{j=0,1} \ketbra{j}^{C''} \otimes (\pi_{\tilde c|z}^{C'})^{T^j} \right) U_{C},
\end{equation}
Here, $T^0$ represents an absence of transposition. Indeed, using lemma \ref{lemma:bowels_selftest} we can replace the auxiliary state and the measurements over Charlie, denoted in general by $C_{\tilde c|z}$, as follows
\begin{equation}
	C_{\tilde c|z}^{C_0} \left(\ket{\psi}^{C_0 B_0} \otimes \ket{00} \otimes \ket{00} \right) = \Pi_{\tilde c|z}^{C} \left(\ket{\psi}^{C_0 B_0} \otimes \ket{00} \otimes \ket{00} \right).
\end{equation}
From this we obtain that the subnormalised state that Bob receives is given by \cite{Bowles_2018_pra} 
\begin{equation}
	\tau_{\tilde c|z} = \tr_{C}\left(C_{\tilde c|z}^{C_0} \left[\ketbra{\psi}^{C_0 B_0} \otimes \ketbra{00}\otimes \ketbra{00} \right]\right) =  \frac{1}{2} U_{ B}^\dagger \left(\sum_{j=0,1} \rho_{\xi_j}^{B_0B''}\otimes (\pi_{\tilde c|z}^{B'})^{T^{j\oplus 1}} \right) U_{ B}.
\end{equation}
Here, $\rho_{\xi_j}^{B_0 B''} = \tr_{C_0}(\ketbra{\xi_j}^{C_0 B_0})\otimes \ketbra{j}^{B''}$ is a subnormalised state.

\subsection{Bowles' self-test for higher dimensions}\label{AppDI_high_dim}
Bowles’ self-test can be generalised to higher dimension via parallelisation. This provides a protocol to self-test the $n$-fold tensor product of the two-qubit maximally entangled state $\ket{\phi^+}^{\otimes n}$  and all combinations of the $n$-fold tensor products of Pauli measurements for Charlie, namely $\{\sigma_Z,\sigma_X,\sigma_Y\}^{\otimes n}$. This is achieved when each of the $n$ parallel Bell tests maximally violates the triple-CHSH inequality in Eq~\eqref{eq:Bell_triple}. We now outline Bowles' parallelised self-test, and refer to Ref~\cite{Bowles_2018_pra} for details.

More specifically, denote by $\vec{z} = (z_1,\dots,z_n)$ and $\vec c = (\tilde c_1,\dots, \tilde c_n)$  Charlie's input vector and output vector. Charlie has three possible input settings $z_i \in \{ 1,2,3\}$ and two possible outcomes $\tilde c_i = \pm 1$. Similarly, denote by $\vec{y} = (y_1,\dots,y_n)$ and $\vec b = (\tilde b_1,\dots, \tilde b_n)$ Bob's input and output vector with $y_i \in \{1,2,3,4,5,6\}$ and $\tilde b_i = \pm 1$. For a fixed value of $i$, we then have three possible inputs for Charlie and six for Bob, corresponding to the self-test in the original setting. For each value of $i$ we then define a Bell operator analogous to \eqref{eq:Bell_triple} for the correlations $\{E_{y,z}^i \}_{y,z}$. Given that $z_i = z$ and $y_i = y$, the average expectation values are
\begin{equation}
	E_{y,z}^i = \frac{1}{3^{n-1}6^{n-1}} \sum_{\substack{\vec z |z_i = z \\ \vec y |y_i = y}} \sum_{\vec b,\vec c} \tilde b_i \tilde c_i\, p(\vec b, \vec c| \vec y, \vec z).
\end{equation}
For each parallel test $i$, we then compute the violation of the triple CHSH-inequality \eqref{eq:Bell_triple} for the correlations $\{E_{y,z}^i \}_{y,z}$. If each Bell test is maximally violated, i.e., $\sum_{i=1}^n \mathcal B_i = n \ 6\sqrt 2$, we can apply the parallel self-testing theorem of Ref~\cite{Bowles_2018_pra} and certify \(n\) maximally entangled qubit pairs and Charlie’s tensor-product Pauli measurements.  In general, however, the $y$-direction of Charlie's local subsystems need not be aligned for different self-tests run in parallel. This comes from the fact that we perform the protocol in lemma \ref{lemma:bowels_selftest} without introducing dependencies between the $n$ individual self-tests. Nevertheless, it was shown in Ref~\cite{Bowles_2018_pra} that it is possible to correlate the flipping of the $\sigma_Y$ measurements, such that either none of them are flipped $\sigma_Y$ or all measurements are flipped $\sigma_Y^T$. Hence, we require that each of the $n$ individual self-test is maximally violated and the flips of the measurements are correlated. Under these conditions and defining $D=2^n$,  Bob receives the subnormalised steered states of the form
\begin{equation}\label{eq:assemblage_highdim}
	\tau_{\vec c| \vec z} = \frac{1}{D} U_{B}^\dagger \left[\sum_{j=0,1} \rho_{\xi_j}^{B_0 B''} \otimes (\pi_{\vec c| \vec z}^{B'} )^{T^{j\oplus 1}}\right] U_{B},
\end{equation}
for a local unitary $U_{B}$  \cite{Bowles_2018_pra}. Here $\oplus$ denotes addition modulo two. The $n$-fold tensor product of Pauli measurements and the $j$-dependent auxiliary state read
\begin{equation}
	\pi_{\vec c| \vec z}^{B'} = \bigotimes_{i=1}^n \pi_{\tilde c_i| z_i}^{B_i'}, \qquad  \rho_{\xi_j} = \tr_{C_0}\left[ \ketbra{\xi_j}^{C_0 B_0}\right] \otimes  \ketbra{j \dots j}^{B''}. 
\end{equation}
Here, $\ket{\xi} = \ket{\xi_0}^{C_0 B_0} \otimes \ket{0\dots0}^{C'' B''}+\ket{\xi_1}^{C_0 B_0} \otimes \ket{1\dots1}^{C'' B''}$, which ensures that flipping of the measurements always is correlated between the $n$ parallel self-tests.

\subsection{Constructing an inequality for certification in the PAB scenario}\label{App_Ineq_Certif}

We now construct a PAB inequality which can be used to certify that a set of preparations $\mathcal E=\{\rho_x\}_x$ does not admit an $r$-complex CCM. This relies on Bowles' parallel self-testing protocol, which lets us certify \(n\) maximally entangled qubit pairs \(|\phi^+\rangle^{\otimes n}\) with local dimension \(D=2^n\). Thus, to deal with states $\rho_x$ of arbitrary dimension \(d\), we embed them into \(\mathbb C^D\) by taking \(n=\lceil\log_2d\rceil\). This embedding indeed preserves classical complexity. Hence, take $D=2^n\ge d$ and regard the $d$-dimensional states as states on $\mathbb C^D$, keeping the notation $\rho_x$. 

To derive a PAB inequality for $r$-complex states, we start by considering the PAB scenario in which Alice emits an $r$-valued message but the broadcaster is quantum and is allowed to distribute a quantum state to Bob and Charlie \cite{Sarubi2026}. This gives correlations of the form
\begin{equation}\label{eq:CQ}
	p(\tilde b,\tilde c|x,y, z)
	=\int d\lambda\,q(\lambda)\sum_{c=1}^r
	p(c|x,\lambda)p_Q(\tilde b,\tilde c|c,y, z,\lambda),
\end{equation}
where $q$ is a probability distribution and the quantum probability distribution admits the decomposition $p_Q(\tilde b,\tilde c|c,y, z,\lambda) = \tr(\rho_{c,\lambda}\left[B_{\tilde b |y,\lambda} \otimes C_{\tilde c|z,\lambda}\right])$, where $\{B_{\tilde b|y,\lambda}\}_{\tilde b}$ and $\{C_{\tilde c|z,\lambda}\}_{\tilde c}$ are Bob's and Charlie's measurements, respectively, and $\rho_{c, \lambda}$ is the bipartite quantum state distributed by the broadcaster. We now apply Bowles' parallel self-test to the correlations in Eq~\eqref{eq:CQ}. If all triple-CHSH Bell tests are maximally violated, the self-test implies that every conditional distribution \(p_Q(\cdot|c,\lambda)\) associated with a nonzero weight $q$ maximally violates the Bell inequalitites. Hence, the self-testing result applies separately to each such \(p_Q(\cdot|c,\lambda)\).

Consider now that Bob has an additional input $y=\star$ with $D^2$ outcomes, which is not used in the self-testing step. Given that self-test was successful, the quantum probabilities takes the form
\begin{equation}\begin{aligned}
p_Q(\tilde b,\vec c|c,y=\star,\vec z,\lambda)=\tr[(\rho_{c,\lambda}\otimes\Psi)(B_{\tilde b|\star}\otimes C_{\vec c|\vec z})]=\tr[(\rho_{c,\lambda}\otimes\tau_{\vec c|\vec z}) B_{\tilde b|\star}] ,
\end{aligned}\label{eq:App_intended}\end{equation}
where $\tau_{\vec c|\vec z}=\tr_C[(\openone\otimes C_{\vec c|\vec z})\Psi]$ is Bob's subnormalized remotely prepared state. Here, we have traced out possible systems on which Charlie acts trivially, meaning that $\rho_{c,\lambda}$ is the additional quantum state that the broadcaster send to Bob. 
Using that the subnormalized states  $\tau_{\vec c|\vec z}$ take the form of Eq~\eqref{eq:assemblage_highdim}, the quantum correlations reads
\begin{equation}
	p_Q(\tilde b,\vec c| c,y=\star,\vec z,\lambda)=\frac1D \tr[G_{\tilde b| c,\lambda}(\pi_{\vec c|\vec z})],
	\label{eq:App_conditional}\end{equation}
where $\pi_{\vec c|\vec z}=\bigotimes_i\pi_{\tilde c_i|z_i}$ is a tensor product of Pauli eigenstates and $\{G_{\tilde b|c,\lambda}\}_{\tilde b}$ is a POVM acting on the $D$-dimensional system. Specifically, the POVMs read 
\begin{equation}
G_{\tilde b|c,\lambda} = \sum_{j=0}^1 \braket{\xi_j} \left(\tilde G_{\tilde b|c,\lambda}^{(j)}\right)^{T^{j\oplus 1}}, \qquad \text{with} \qquad  \tilde{G}_{\tilde b|c,\lambda}^{(j)}= \tr_A\left((\rho_{c\lambda}^A \otimes \openone^{B'})\tilde B_{\tilde b|\star,\lambda}^{(j)} \right)
\end{equation}
and the joint measurement reads
\begin{equation}\label{eq:M_j}
	\tilde B_{\tilde b|\star,\lambda}^{(j)} = \tr_{B_0 B''}\left( \left[ \openone^{AB'} \otimes \tilde \rho_{\xi_j} ^{B_0 B''} \right] \left[ (\openone^A \otimes U_B) B_{\tilde b | \star,\lambda}^{AB} (\openone^A \otimes U_B^\dagger) \right] \right).
\end{equation}
Here we have intoduced the notation $\rho_{\xi_j}  =  \braket{\xi_j} \tilde \rho_{\xi_j} $. Since $\tilde G_{\tilde b|c,\lambda}^{(j)}$ is a POVM and transposition preserves positivity and normalization it follows that $G_{\tilde b|c,\lambda}$ is a POVM. Substitution into \eqref{eq:CQ} yields
\begin{equation}
	p(\tilde b,\vec c|x,y=\star,\vec z)=\frac1D\tr[M_{\tilde b|x}\pi_{\vec c|\vec z}],\quad \text{with} \quad M_{\tilde b|x}=\int d\lambda\,q(\lambda)\sum_{c=1}^r p(c|x,\lambda) G_{\tilde b|c,\lambda}.
	\label{eq:App_effective}
\end{equation}
Thus, the effective measurements admit the classical MDI decomposition \eqref{eq:M_classical}. Since the set of such effective families is compact and convex, any set $\mathcal M=\{M_{\tilde b|x}\}$ can be detected via a separating witness 
\begin{equation}
	\mathcal W_r(\mathcal M)=\sum_{\tilde b,x}\tr(\mathcal O_{\tilde b,x}M_{\tilde b|x})\le0,
	\label{eq:App_separation}\end{equation}
where the bound holds for every classically decomposable $\mathcal M$. Next, using that the tensor products of Pauli eigenstates $\pi_{\vec{c}|\vec{z}}$ are informationally complete, the Hermitian witness operators admit an expansion
\[
\mathcal O_{\tilde b,x}=\sum_{\vec c,\vec z}a_{\tilde b\vec c x\vec z}\pi_{\vec c|\vec z}
\]
with real coefficients. This gives the inequality
\begin{equation}
	\mathcal W_r=\sum_{\tilde b,\vec c,x,\vec z}a_{\tilde b\vec c x\vec z}
	p(\tilde b,\vec c|x,y=\star,\vec z)\le0.
	\label{eq:App_witness}\end{equation}
The bound follows from \eqref{eq:App_effective} for classical PAB models satisfying the testing conditions. Consequently, a positive witness value together with the maximal-score and alignment observations excludes every $r$-message classical PAB model. The following subsection establishes an experiment for which this occurs whenever the states admit no $r$-message CCM.



\subsection{Certifying the failure of CCMs in the PAB scenario}
We now describe how to certify that a set of $d$-dimensional preparations $\mathcal E = \{\rho_x\}_x$ is not $r$-complex.  As in the previous subsections, we embed the states into dimension $D=2^n\ge d$. Consider that each of the $n$ parallel self-tests was successful, meaning that each Bell test reached a maximal violation of the associated triple CHSH inequality, i.e., $\mathcal B_i = 6 \sqrt 2$ for $i = 1,\dots,n$, and that the additional observations certifying the correlated transpositions are satisfied for every $x$.  
Let the auxiliary state to be $\Psi = \ketbra{\phi^+}$ and take Charlie's measurement to be a informational complete set of tensor products of Pauli measurements.   Let Bob’s additional measurement $y = \star$ to be the projection onto the maximally entangled basis $(\openone \otimes U_{\tilde b}) \ket{\phi^+}$, where $\{U_{\tilde b}\}_{\tilde b=1}^{D^2}$ constitute a Hilbert-Schmidt orthogonal unitary basis. The corresponding correlations are
\begin{equation}
	p(\tilde b, \vec c|x,y = \star, \vec z) = \tr(\left[\rho_x \otimes \ketbra*{\phi^+} \right] \left[(\openone \otimes U_{\tilde b}) \ketbra*{\phi^+}(\openone \otimes U_{\tilde b}^\dagger) \otimes \pi_{\vec c| \vec z} \right]) = \frac{1}{D} \tr(M_{\tilde b|x} \pi_{\vec c| \vec z}),
\end{equation} 
where we have identified $M_{\tilde b|x} = \frac{1}{D}  U_{\tilde b}^* \rho_x U_{\tilde b}^T$. We now follow the same line of reasoning as in section \ref{secMDI} in the main text. If $M_{\tilde b|x}$ admits an MDI decomposition, then
\begin{equation}
	\frac{1}{D}  U_{\tilde b}^* \rho_x U_{\tilde b}^T =  \int d\lambda q(\lambda) \sum_{c=1}^r p(c|x,\lambda)  G_{\tilde b|c,\lambda}.
\end{equation}
Solving this equation for $\rho_x$ and averaging both sides of the resulting expression over the possible outcomes $\tilde b$ yields
\begin{equation}
	\rho_x = \int d\lambda q(\lambda) \sum_{c=1}^r p(c|x,\lambda)  \underbrace{ \frac{1}{D}\sum_{\tilde b = 1}^{D^2} \left(U_{\tilde b}^T G_{\tilde b|c,\lambda} U_{\tilde b}^* \right)}_{=\sigma_{c,\lambda}} =  \int d\lambda q(\lambda) \sum_{c=1}^r p(c|x,\lambda) \sigma_{c,\lambda} .
\end{equation}
Since $\sigma_{c,\lambda}$ is a state, this leads to an $r$-complex model for the embedded states and therefore an $r$-complex CCM for the original dimension $d$. By contrapositive, if $\mathcal E$ admits no $r$-complex CCM, the family $M_{\tilde b|x}$ lies outside the classical MDI set. The witness argument of subsection \ref{App_Ineq_Certif} therefore provides coefficients $a_{\tilde b\vec c x\vec z}$, chosen for this family, such that
\begin{equation}
	\mathcal W_r= \sum_{\tilde b,\vec c,x,\vec z}a_{\tilde b\vec c x\vec z}  p(\tilde b,\vec c|x,y=\star,\vec z) = \frac1D\sum_{\tilde b,\vec c,x,\vec z}a_{\tilde b\vec c x\vec z}
	\tr[M_{\tilde b|x}\pi_{\vec c|\vec z}]>0.
\end{equation}
Together with the testing conditions, this violation excludes every $r$-message classical PAB model, and certifies the failure of an $r$-complex CCM for $\mathcal E$.

\section{Classical communication models for pure quantum states with isotropic noise}\label{AppClassicalModels}
In this section we first derive the optimal classical communication model for simulating the transmission of all quantum states under isotropic noise using an $r$-valued classical message. Thereafter, we derive the lower and upper bounds on the critical visibility presented in Result~\ref{ResultCriticalVisibility}. 

Consider all pure $d$-dimensional states when subject to isotropic noise $\mathcal E = \{\rho_\psi\}_\psi$ where $\rho_\psi = v \ketbra{\psi} + \frac{1-v}{d} \openone$. Let $v^*(\mathcal E)$ be the critical visibility, below which the set $\mathcal E$ has an $r$-complex CCM and above which it does not. The average overlap between $\rho_\psi$ and its associated pure state $\ket{\psi}$ is given by
\begin{equation}
\int d\psi \ \langle \psi|\rho_\psi|\psi\rangle = \frac{v(d-1)+1}{d}
\end{equation}

Assume that $v \leq v^*(\mathcal E)$, such that the set $\mathcal E$ is $r$-complex. Then each noisy state can be decomposed as $\rho_{\psi} = \int d\lambda q(\lambda) \sum_{c=1}^r p(c|\psi,\lambda) \sigma_{c,\lambda}$. The optimal average overlap can then be expressed as 
\begin{equation}\label{eq:int}
 \int d\psi \ \langle \psi|\rho_\psi|\psi\rangle \leq \max_{\lambda} \int d\psi \sum_{c=1}^r p(c|\psi,\lambda) \langle \psi |\sigma_{c,\lambda} | \psi\rangle  \leq  \max_\lambda \int d\psi \max_{c}\langle \psi |\sigma_{c,\lambda} | \psi\rangle  \leq \max_{\phi_1,\dots,\phi_r \in \mathbb{C}^d} \int d\psi \max_{c} |\langle \psi |\phi_c \rangle|^2 \equiv \mathcal I_{d,r} 
\end{equation}
Here we use that Bob's preparations, associated with the optimal choice of $\lambda$, can optimally be taken to be pure i.e., $\sigma_{c,\lambda} = \ketbra{\phi_c}$. Hence, the ensemble $\mathcal E$ violates the classical bound if and only if 
\begin{equation}\label{eq:crit_vis}
v > v_{\text{witness}} \equiv  \frac{d\mathcal I_{d,r}-1}{d-1}.
\end{equation}

To prove that $v_{\text{witness}}$ is equivalent to the critical visibility we construct an explicit CCM. To this end, denote Bob's optimal set of preparations in Eq~\eqref{eq:int} by 
\begin{equation}
\{\phi_c^*\}_{c=1}^r = \arg \max_{\phi_1,\dots,\phi_r \in \mathbb{C}^d} \int d\psi \max_{c} |\langle \psi |\phi_c \rangle|^2.
\end{equation}
Now, let Alice and Bob share unlimited classical randomness $\lambda$. To each variable $\lambda$ associate a unique unitary $U \in U(d)$, such that there is a one-to-one mapping $\lambda \leftrightarrow U$. Take Alice's encoding function to be deterministic $D(c|\psi, U)$, such that she deterministically encode each state $\ket{\psi}$ into a classical message $c \in \{1,\dots,r\}$ conditioned on the hidden variable. Upon receiving the message, Bob applies a completely positive trace-preserving map $\Lambda$ to the classical state $\ketbra{c}$, and then rotates the state with the unitary $U$ using his knowledge of the shared variable. This results in Bob's output state
\begin{equation}
\rho_\psi = \int d\mu(U) \sum_{c=1}^r D(c|\psi, U) U\Lambda(\ketbra{c}) U^\dagger,
\end{equation}
Specifically, take Bob's quantum channel to be $\Lambda(\cdot) = \sum_{c=1}^r \langle c|\cdot|c\rangle \ketbra{\phi_c^*}$. Moreover, take Alice's encoding as $D(c|\psi, U) = 1$ for $c = \arg \max_{1\leq k \leq r}\left|\langle \psi| U | \phi_k^*\rangle \right|^2$ and zero otherwise. This model is invariant under the stabilizer group of $\psi$, i.e.  under all unitary transformations leaving invariant $\psi$. Hence,  it yields a state of the form $\rho_\psi = v \ketbra{\psi} + \frac{1-v}{d} \openone$.  It is left to compute $v$, which can be obtained by the overlap $\expval{\rho_\psi}{\psi}$:
\begin{equation}\label{eq:integral}
\frac{v(d-1)+1}{d} =  \langle \psi|\rho_\psi|\psi\rangle =  \int d\mu(U) \sum_{c=1}^r D(c|\psi, U) \left|\langle \psi| U | \phi_c^*\rangle \right|^2 =  \max_{\phi_1,\dots,\phi_r \in \mathbb{C}^d} \int d \psi \max_{c} \left|\braket{\psi}{\phi_c}\right|^2 = \mathcal I_{d,r}.
\end{equation}
In the second step we have identified the integration over unitary transformations with that over pure states. 
Consequently, an $r$-complex classical communication model can simulate the transmission of all states for any visibility
\begin{equation}\label{eq:crit_vis}
v \leq v_{\text{model}} \equiv \frac{d\mathcal I_{d,r}-1}{d-1}.
\end{equation}
Since $v^* = v_{\text{model}}=v_{\text{witness}}$ the critical visibility above which no CCM exists is exactly given by $v^*$. In general, we do not know how to solve the integral. Therefore, we instead derive upper and lower bounds below when $r>d$ and $d \geq 2$.

\subsection{Lower bound on the critical visibility}
We now compute a lower bound on the expression $\mathcal I_{d,r}$ in \eqref{eq:integral} for an $r$-complex communication model. To this end, we omit the optimisation over Bob's preparations $\{\ket{\phi_c}\}_{c=1}^r$ and instead take them to be Haar random. This results in that the states $\phi_c$ are independent variables and gives the lower bound
\begin{equation}
\mathcal I_{d,r} \geq \mathcal I_L \equiv \int d\psi  \int d\phi_1 \dots \int d\phi_r \left( \max_{1\leq c \leq r}|\braket{\psi}{\phi_c}|^2 \right).
\end{equation}
For fixed $\psi$, we next use that the integrals of the function $f_\psi(\phi_1,\dots,\phi_r)=  \max_{1\leq c \leq r} |\braket{\psi}{\phi_c}|^2$ can be expressed as the expected value over Haar measures $\mu_{\text{H}}$ \cite{Mele_2024}. This yields
\begin{equation}
 \mathcal I_L =  \int d\psi \left(\underset{\phi_1,\dots, \phi_r \sim \mu_{\text{H}}}{ \mathbb{E}}\left[\max_c |\braket{\psi}{\phi_c}|^2  \right] \right) .
\end{equation}
To further simplify the expression we need the layer-cake identity, which states that the expectation value over the non-negative random variable $X$ is the area under the tail probability curve \cite{expectation_value}
\begin{equation*} 
	\mathbb{E}[X] = \int_0^\infty dt \mathbb{P}(X > t) = \int_0^\infty dt (1- \mathbb{P}(X \leq t)).
\end{equation*}
We now employ this identity to the lower bound. Since the joint probability that the function $f_\psi(\phi_1,\dots,\phi_r) > 1$ is zero, this gives that
\begin{equation}
 \mathcal I_L =   1-  \int d\psi  \int_0^1 dt \  \underset{\phi_1,\dots, \phi_r \sim \mu_{\text{H}}}{ \mathbb{P}} \left[ \max_{1\leq c \leq r} |\braket{\psi}{\phi_c}|^2 \leq t \right]
\end{equation}
For fixed $\psi$, the condition $|\braket{\psi}{\phi_c}|^2 \leq t$ must hold for all independent Haar random pure states $\phi_c$. This implies that the joint probability factorises over the Haar measures. 
\begin{equation}
 \mathcal I_L =   1-   \int_0^1 dt   \int d\psi   \prod_{c = 1}^r \underset{\phi_c \sim \mu_{\text{H}}}{\mathbb{P}} \left[|\braket{\psi}{\phi_c}|^2 \leq t \right]
\end{equation}
For fixed unit vector $\psi \in \mathbb {C}^d$, the overlap between the Haar distributed state $\phi_c$ and $\psi$ follows a Beta distribution, i.e. ,  $|\braket{\psi}{\phi_c}|^2 \sim Beta(1,d-1)$ \cite{Bengtsson_2017}. Thus, each factor in the product can be expressed as
\begin{equation}\label{eq:beta}
	\underset{\phi_c \sim \mu_{\text{H}}}{ \mathbb{P}} \left[|\braket{\psi}{\phi_c}|^2 \leq t \right] = \frac{1}{B(1,d-1)}\int_0^t dx \ (1-x)^{d-2} = (d-1)\left[\frac{(1-x)^{d-1}}{d-1} \right]_t^0 = 1-(1-t)^{d-1},
\end{equation}
where $B(\alpha,\beta) = \frac{\Gamma(\alpha)\Gamma(\beta)}{\Gamma(\alpha+\beta)}$ and $\Gamma$ is the Gamma-function with $\alpha = 1$ and $\beta = d-1$. Note that Eq~\eqref{eq:beta} is independent of $\psi$. This results in that
\begin{equation}\label{eq:lb_final}
\mathcal I_L = 1- \int_0^1 dt \ (1-(1-t)^{d-1})^r = 1-  \frac{\Gamma( \frac{1}{d-1} +1 )\Gamma(r+1)}{\Gamma( \frac{1}{d-1}+r+1)}.
\end{equation}
Using the Gamma function properties that $\Gamma(r+1) = r!$ and $\Gamma(z+1) = z$, we can simplify the last fraction as follows
\begin{equation}
P_{d,r} \equiv \frac{\Gamma( \frac{1}{d-1} +1 )\Gamma(r+1)}{\Gamma( \frac{1}{d-1}+r+1)} = \frac{ \Gamma( \frac{1}{d-1} +1 ) r!}{\Gamma( \frac{1}{d-1} +1 ) \prod_{c = 1}^r \left(\frac{1}{d-1}+c\right)} = \prod_{c=1}^r \frac{c}{\frac{1}{d-1}+c}
\end{equation}
Inserting the expression of $\mathcal I_L$ into Eq~\eqref{eq:crit_vis} we obtain a lower bound for the critical visibility
\begin{equation}
v^*(\mathcal E)  \geq 1- \frac{d}{d-1} \prod_{c=1}^r \frac{c}{\frac{1}{d-1}+c} = 1-  \prod_{c=2}^r \frac{c}{\frac{1}{d-1}+c}.
\end{equation}
Below this visibility all sets of $d$-dimensional states admit an $r$-complex classical communication model.

\subsubsection{Asymptotic behaviour}
We now comment on the asymptotic behaviour of the lower bound $\mathcal I_L$ in Eq~\eqref{eq:lb_final}. Start by expressing the lower bound as $\mathcal I_L = 1- P_{d,r}$ where we have introduced the notation
\begin{equation}
P_{d,r} \equiv \prod_{c=1}^r \frac{1}{ 1 + \frac{a}{c}}, \qquad a = \frac{1}{d-1}.
\end{equation}
Let us first consider the case $d=2$, for which $a=1$ and hence
\begin{equation}
	P_{2,r}=\frac{1}{1+r}\implies \mathcal I_L=1-\frac{1}{r+1},
\end{equation}
which trivially scales as $1-\Gamma (a+1) r^{-a}=1-r^{-1}$. Let us move to the $d>2$ case. Computing the logarithm of $P_{d,r}$ we find that
\begin{equation}
\log(P_{d,r}) = -\sum_{c=1}^r \log(1+\frac{a}{c}) =  \sum_{c=1}^r \sum_{n=1}^{\infty} \frac{1}{n} \left(\frac{-a}{c}\right)^n = \sum_{n=1}^{\infty} \frac{(-a)^n}{n}  \sum_{c=1}^r \frac{1}{c^n}  = -a \sum_{c=1}^r \frac{1}{c} + \sum_{n=2}^\infty \frac{(-a)^n}{n} \sum_{c=1}^r \frac{1}{c^n}.
\end{equation}
In the second step we have used the Taylor expansion of $\log(1+x) = -\sum_{n=1}^\infty \frac{(-x)^n}{n}$ for $|x|<1$. The first term in the last step, associated with $n = 1$, corresponds to the standard Harmonic series with $r$ terms, denoted by $H_r$, while the second term, associated with $n>1$, can be expressed in terms of the Riemann zeta-function $\zeta(n) = \sum_{c =1}^\infty \frac{1}{c^n}$. This gives that
\begin{equation}
\log(P_{d,r}) =  -a H_r+ \sum_{n=2}^\infty \frac{(-a)^n}{n} \left(\zeta(n) - \sum_{c=r+1}^\infty \frac{1}{c^n}\right) = -a H_r+ \sum_{n=2}^\infty \frac{(-a)^n}{n} \zeta(n) - \sum_{n=2}^\infty  \frac{(-a)^n}{n} \sum_{c=r+1}^\infty \frac{1}{c^n}
\end{equation}
Next, we use the Taylor series of $\log(\Gamma(1+a))$ with $|a| < 1$, namely
\begin{equation}
	\log(\Gamma(1+a)) = - \gamma a + \sum_{k=2}^\infty \frac{(-a)^k}{k}\zeta(k)
\end{equation}
where $\gamma$ is the Euler–Mascheroni constant. Thus,
\begin{equation}\label{eq:log_P}
\log(P_{d,r}) = -a H_r + a \gamma + \log(\Gamma(1+a)) - \sum_{n=2}^\infty \frac{(-a)^n}{n} \sum_{c=r+1}^\infty \frac{1}{c^n}.
\end{equation}

We now study the limit in which $r\gg 1$ for fixed $d$. To this end, the first $r$ terms in the Harmonic series can be approximated as $H_r \approx \log(r) + \gamma$. To simplify the last term, we note that $f(c) = c^{-n}$ is a positive function that decreases monotonically for $n > 1$. Hence, 
\begin{equation}
	f(c) \leq \int_{c-1}^{c} f(c') dc'
\end{equation}
Consequently, this leads to 
\begin{equation}
	  \sum_{c=r+1}^\infty f(c) \leq \sum_{c=r+1}^\infty  \int_{c-1}^{c} f(c') dc' =  \lim_{M \rightarrow \infty} \int_{r}^{M} \frac{1}{c^n} dc = \lim_{M \rightarrow \infty} \frac{1}{n-1}\left(\frac{1}{r^{n-1}}-\frac{1}{M^{n-1}}\right) = \frac{r^{-(n-1)}}{n-1}
\end{equation}
Hence, the largest contribution in the sum $ \sum_{n=2}^\infty \frac{(-a)^n}{n} \frac{r^{1-n}}{n-1}$ occurs for $n = 2$, which is of order $\mathcal O(r^{-1})$. Hence, in the limit of large $r$, we have that
\begin{equation}\label{eq:log_P}
\log(P_{d,r}) \approx -a \log(r) +  \log(\Gamma(1+a))  +  O(r^{-1}).
\end{equation}
Consequently, the lower bound grows as
\begin{equation}\label{eq:approx_lb}
\mathcal I_L = 1- \prod_{k=1}^r \frac{k}{ \frac{1}{d-1}+k} \approx 1- \exp[ \log(r^{-a}) + \log(\Gamma(1+a)) ] = 1- \Gamma(a+1)r^{-a}
\end{equation}

\subsection{Upper bound on the critical visibility}
We now compute upper bounds on the expression $\mathcal I_{d,r}$. 

\subsubsection{When $r \leq d$}
We first consider the case in which $r \leq d$. Denote by $\{\phi_c^*\}_c$ the $r$ unit vectors that optimise the integral $\mathcal{I}_{d,r}$. Next, we use that any Haar random state $\psi$ can be obtained by normalising an unnormalised pure state $\ket*{\tilde \psi}$ using independent complex Gaussian coefficients in some orthogonal basis $\{\ket{1},\dots,\ket{d}\}$ \cite{Jones_2007}
\begin{equation*}\
	 \ket*{\tilde \psi} = \sum_{j =1}^{d} z_j \ket{j}.
\end{equation*}
Here, $z_j$ are zero-mean Gaussian variables with the properties $\langle z_j^* z_k \rangle = \delta_{j,k}$ and $\langle z_k z_k \rangle =0$ . Let each variable $z_j$ have the same marginal density $f_j(z) = \frac{1}{\pi}e^{-|z|^2}$. Writing $ \ket*{\tilde \psi} = m \sqrt{d} \ket{\psi}$, we denote the measure of this set by $d \mu_G (\psi,m)$. Due to unitary invariance of the Haar measure, this set factorises $d \mu_G (\psi,m) = d \psi d \mu_G(m)$ where the normalisation term of the weights reads $ \int d \mu_G(m) m^2 = 1$. Thus, it follows that
\begin{equation}
	\mathcal{I}_{d,r} =  \max_{\phi_1,\dots,\phi_r  \in \mathbb C^d}  \int d\psi \max_{c} \left|\braket{\phi_c}{\psi}\right|^2 = \frac {1}{d}  \int d \mu_G (\psi,m) \max_{c} \lvert\braket*{\phi^*_c}{\tilde \psi}\rvert^2 =  \frac {1}{d} \left( \int_0^{\infty} dt 1-\underset{\tilde \psi \sim \mu_G}{\mathbb{P}}\left[ \max_c\lvert\braket*{\phi^*_c}{\tilde \psi}\rvert^2 \leq t \right]\right)
\end{equation}
Since $\phi_c^*$ do not need to be orthogonal, the overlaps $ \lvert\braket*{\phi^*_c}{\tilde \psi}\rvert$ are in general not independent. However, for every unit vector $\phi_c^*$ the overlap $g_c = \braket*{\phi^*_c}{\tilde \psi}$ corresponds to a standard complex Gaussian variable and has a complex standard Gaussian distribution with probability density $\frac{1}{\pi}e^{-|z|^2}$.  This means that the joint probability that $|g_c| \leq \sqrt{t}$ for all $c = 1,\dots,r$ is greater than or equal to the probability that $r$ independent standard complex Gaussian variables simultaneously fulfil $|g_c| \leq \sqrt{t}$ \cite{Lata_a_2017}. Hence, we have that 
\begin{equation}
\underset{\tilde \psi \sim \mu_G}{\mathbb{P}}\left[ \max_c \lvert\braket*{\phi^*_c}{\tilde \psi}\rvert \leq \sqrt{ t} \right] = \underset{\tilde \psi \sim \mu_G}{\mathbb{P}}\left[ \lvert\braket*{\phi^*_1}{\tilde \psi}\rvert \leq \sqrt{t}, \dots, \lvert\braket*{\phi^*_r}{\tilde \psi}\rvert \leq \sqrt{t} \right] \geq \prod_{c = 1}^r \underset{\tilde \psi \sim \mu_G}{\mathbb{P}}\left[ \lvert\braket*{\phi^*_c}{\tilde \psi}\rvert \leq \sqrt{t} \right] 
\end{equation}
In polar coordinates this yields
\begin{equation}
\underset{\tilde \psi \sim \mu_G}{\mathbb{P}}\left[ \lvert\braket*{\phi^*_c}{\tilde \psi}\rvert \leq \sqrt{t} \right] = \int_0^{2 \pi} \int_0^{\sqrt{t}} \frac{1}{\pi}e^{-r^2} r dr d\theta = 1- e^{-t}.
\end{equation}
Since this is independent of $\phi_c^*$, we obtain the upper bound
\begin{equation}
	\mathcal{I}_{d,r} \leq  \frac {1}{d} \left( \int_0^{\infty} dt 1-(1- e^{-t})^r\right) = \frac{H_r}{d}.
\end{equation}
Hence, for $r\leq d$ it holds that $\mathcal{I}_{d,r} \leq \frac{H_r}{d}$. Moreover, this bound is attained by taking all $\phi_c$ to be orthogonal, e.g. $\phi_c = \ket{c}$.

\subsubsection{For $r>d$}
Lastly,  we consider the case $r>d$.  We start by rewriting the integral in terms of the expected value,  so that
\begin{equation}
\mathcal I_{d,r}=  \max_{\phi_1,\dots,\phi_r  \in \mathbb C^d}\underset{\psi \sim \mu_{\text{H}}}{ \mathbb{E}}\left[\max_c |\braket{\psi}{\phi_c}|^2\right]=
\max_{\phi_1,\dots,\phi_r  \in \mathbb C^d}  \int_0^1 dt \underset{\psi \sim \mu_{\text{H}}}{ \mathbb{P}}\left[\max_c |\braket{\psi}{\phi_c}|^2 > t  \right].
\end{equation}
In order to get an upper bound, we use a union-bound over the probability distribution of $\max_c \left|\braket{\psi}{\phi_c}\right|^2$:
\begin{equation}
 \underset{\psi \sim \mu_{\text{H}}}{ \mathbb{P}}\left(\max_c \left|\braket{\psi}{\phi_c}\right|^2 > t\right)= \underset{\psi \sim \mu_{\text{H}}}{ \mathbb{P}}\left( \left|\braket{\psi}{\phi_1}\right|^2 > t \vee \cdots \vee\left|\braket{\psi}{\phi_r}\right|^2 > t\right)\leq \sum_{c=1}^r  \underset{\psi \sim \mu_{\text{H}}}{ \mathbb{P}}\left( \left|\braket{\psi}{\phi_c}\right|^2 > t\right)=r(1-t)^{d-1}.
\end{equation}
In the last step, we have identified that the overlap $\left|\braket{\psi}{\phi_c}\right|^2 \sim Beta(1,d-1)$ is Beta distributed for fixed $\phi_c$ and Haar distributed $\psi$, meaning that each term in the sum is independent on $\phi_c$ and subsequently allows us to use Eq~\eqref{eq:beta}. Since in addition the probability has to be lower than one, we get
\begin{equation}
 \underset{\psi \sim \mu_{\text{H}}}{ \mathbb{P}}\left(\max_c \left|\braket{\psi}{\phi_c}\right|^2 > t\right)\le \min\{1,r(1-t)^{d-1}\}.
\end{equation}
The bound is independent on the choice of $\phi_c$,  so it also applies when considering the set maximizer, which leads to
\begin{equation}
\mathcal I_{d,r}\le \mathcal I_U \equiv \int_0^1 dt \ \min\{1, r(1-t)^{d-1}\} = \int_0^{t^*} dt + \int_{t^*}^1 dt \  r(1-t)^{d-1} = 1- \frac{d-1}{d} r^{-\frac{1}{d-1}} ,
\end{equation}
where the critical value below which $ r(1-t)^{d-1} > 1$ is given by $t^*= 1-r^{-\frac{1}{d-1}}$. This gives an upper bound on the critical visibility
\begin{equation}
v^*(\mathcal E) \leq 1 - r^{-\frac{1}{d-1}}.
\end{equation}

\subsection{Combined integral bounds and asymptotic behaviour}
Finally, we combine the bounds derived above with the result obtained in Ref~\cite{Cobucci2026}. To this end, we find that for $r\ge d$
\begin{equation}
\max\{1- \prod_{k=1}^r \frac{k}{ \frac{1}{d-1}+k}, \frac{H_d}{d}\} \leq \mathcal I_{d,r} \leq 1- \frac{d-1}{d} r^{-\frac{1}{d-1}},
\end{equation}
where $H_d = \sum_{k=1}^d 1/k$ is the Harmonic sum. Using the approximative expression for the lower bound in Eq~\eqref{eq:approx_lb}, valid for $r \gg 1$, we find that the first lower bound overtakes the second, approximately for
\begin{equation}
r = \left(\frac{d-H_d}{d \Gamma(1+\frac{1}{d-1})}\right)^{-(d-1)}.
\end{equation}
The critical visibility below which a CCM exists, and above which it does not, reads
\begin{equation}
 \max\{1- \frac{d}{d-1} \prod_{k=1}^r \frac{k}{\frac{1}{d-1}+k},\frac{H_d-1}{d-1}\}  \leq v^*(\mathcal E) \leq 1 - r^{-\frac{1}{d-1}}.
\end{equation}

\section{Proof of ETF witness bound}\label{App:ETF}
In this appendix, we prove the upper bound for the ETF witness. Thus, consider the following witness 
\begin{equation}
W_\mathrm{ETF} = \sum_{a=1}^n \langle f_a| \sigma_a | f_a \rangle.
\end{equation}
where the set $\{\ket{f_a}\}_a$ corresponds to a $d$-dimensional equiangular tight frame, i.e., $|\braket{f_a}{f_{a'}}|^2 = \frac{n-d}{d(n-1)}$ for $a\neq a'$. Next, assume that the set of $d$-dimensional states $\mathcal E = \{\sigma_a\}_{a=1}^n $ admits an $r$-complex CCM. Then the witness is upper bounded by
\begin{equation}
W_\mathrm{ETF}  \leq  \max_{\gamma} \sum_{c = 1}^r \lambda_{\max} \left(\underbrace{\sum_{a=1}^n D_\gamma(c|a) \ketbra{f_a}}_{= O_c}\right)
\end{equation}
where $\lambda_{\max}$ denotes the largest eigenvalue of the Hermitian operator $O_c \succeq 0$ with eigenspectrum $\mu_{c,1} \geq \mu_{c,2} \geq ... \geq \mu_{c,d} \geq 0$. We now aim to bound the value of the largest eigenvalue $\lambda_{\max}(O_c) \equiv \mu_{c,1}$. To this end we use that
\begin{equation}\label{eq:Oc}
\frac{(\tr( O_c)- \mu_{c,1})^2}{d-1} \leq \sum_{i=2}^d \mu_{c,i}^2 = \tr(O_c^2)- \mu_{c,1}^2,
\end{equation}
where the lower bound is obtained using Cauchy-Schwarz inequality.  Solving for $\mu_{c,1}$ yields the inequality
\begin{equation}\label{eq:ineq}
 \mu_{c,1} \leq \frac{1}{d}\tr(O_c) + \frac{1}{d}\sqrt{(d-1)(d\tr(O_c^2)-\tr(O_c)^2)},
\end{equation}
Denote the total trace of the operator $\mathcal O_c$ by
\begin{equation}
k_c = \tr(\mathcal O_c) = \sum_{a=1}^n D_\gamma(c|a).
\end{equation}
From this, we obtain that the second-order moments reads
\begin{equation}
\tr(\mathcal O_c^2) = k_c + \frac{n-d}{d(n-1)} \sum_{a \neq a'} D_\gamma(c|a) D_\gamma(c|a') = k_c\left(1+\frac{n-d}{d(n-1)}(k_c-1)\right).
\end{equation}
Inserting this back into the inequality \eqref{eq:ineq}, we obtain that
\begin{equation}
\mu_{c,1} \leq  \frac{1}{d}\left(k_c + (d-1) \sqrt{\frac{k_c}{n-1}(n-k_c)}\right) \equiv g(k_c).
\end{equation}
Moreover, since $\frac{\partial^2 g(k_c) }{\partial k_c^2} \leq 0$ for $k_c \in [0,n]$, the largest eigenvalue $\mu_{c,1}$ is upper bounded by the concave function $g$. Thus, Jensen's inequality implies that 
\begin{equation}
W_\mathrm{ETF} \leq   \max_{\gamma} \sum_{c = 1}^r g(k_c) \leq r g\left(\frac{1}{r}\sum_{c=1}^r k_c\right) = r g\left(\frac{n}{r}\right).
\end{equation}
Simplifying the expression we find
\begin{equation}
{W}_\mathrm{ETF}  \leq \frac{n}{d}\left(1 + (d-1)\sqrt{\frac{r-1}{n-1}} \right).
\end{equation}

\section{Algorithmic methods for classical complexity}\label{App:enhancing_SDP}
In this appendix we discuss how to implement the inner and outer approximations of classical complexity. 

\subsection{Inner approximation: constructing the pool $\mathcal A$}
We describe how we build the large pool of candidate strategies $\mathcal A$ used for the inner approximation, where each candidate corresponds to a tuple of $r$ Bloch vectors $\mathcal U_\lambda = (u_{1,\lambda},\dots,u_{r,\lambda})$. The candidates were constructed using several different methods.

\subsubsection{Structured candidates}
We first include a set of structured candidates. The aim is that these tuples combined approximately cover the Bloch sphere. Thus, we first generate a fiducial tuple $\mathcal U_{\text{fid}}=(\vec u_{1},\dots,\vec u_{r})$ for which the $r$ Bloch vectors are approximately uniformly spread over the Bloch sphere. For messages of size $r \in \{2,3,4,6,8,12,20\}$ we hard-code $\mathcal U_{\text{fid}}$ to form an antipodal pair, trine, tetrahedron, octahedron, cube, icosahedron and dodecahedron in the Bloch sphere, respectively. Otherwise, the $r$ Bloch-vectors are taken as the Fibonacci lattice. In this case, we take 
\begin{equation}
	\vec u_{c} \leftarrow (\sqrt{1-z_c^2}\cos(\phi_c),\sqrt{1-z_c^2}\sin(\phi_c),z_c),
\end{equation}
where $z_c = 1-(2c-1)/r$ and $\phi_c = (c-1) \varphi$ and $\varphi = (3- \sqrt{5}) \pi$ is the golden angle. This "golden-angle spiral" yields quasi-uniform point spacing. We then sample a few hundred random unitaries $V^{(m)}$ and construct a sequence of candidates $\mathcal U_m = V^{(m)} \mathcal U_{\text{fid}}$. For sufficiently large $m$ we get an approximately uniform set of rotations over $\mathcal U_{\text{fid}}$.

\subsubsection{Data adapted candidates 1}
We now construct candidates that are adapted to the set of quantum preparations $\{\vec n_x\}_x$ that we aim to simulate. Thus, we want to find a tuple $\mathcal U$, such that each target vector $\vec n_x$ has a large overlap with one of $r$ vectors $\vec u_c \in \mathcal U$. To this end, split the $n$ Bloch vectors $(\vec n_1,\dots,\vec n_n)$ into $r$ groups $G_1,\dots,G_r$. Given that $c^*= \arg \max_{1 \leq c \leq r} \langle \vec n_x,\vec u_c\rangle$ we take $\vec n_x \in G_{c^*}$. The sum of the overlaps between the target state and their closest approximations reads
\begin{equation}\label{eq:Lloyd}
	\mathcal S = \sum_{x=1}^n \max_c \langle \vec n_x, \vec u_c\rangle = \sum_{c=1}^r  \sum_{x \in G_c}  \langle \vec n_x, \vec u_c\rangle \leq \sum_{c=1}^r   \| \sum_{x \in G_c}   \vec n_x\| .
\end{equation}
The upper bound is reached by choosing the unit vector $\vec u_c$ to point along the sum $\sum_{x \in G_c}\vec n_x$. There are $S(n,r)$ \eqref{eq:stirling} different ways we can partition $n$ elements into $r$ non-empty groups. However, since we only seek a good but not optimal splitting of the elements we use a so-called Lloyd clustering method \cite{clustering}. Below we outline the main steps in the clustering algorithm
\begin{enumerate}
	\item Pick $r$ distinct unit vectors $\vec u_1,\dots, \vec u_r$ among the normalised target vectors $\hat n_1,\dots \hat n_n$ where $\hat n_x = \vec n_x /\|\vec n_x\|$. We choose the initial vectors such that they are spread out among the target vectors.
	\begin{enumerate} 
		\item Assign $\vec u_1 \leftarrow \hat n_x$ with probability proportional to $\|\vec n_x\|$ and add the index $x$ to a list $T$.
		\item Assign $\vec u_2 \leftarrow \hat n_x$ with probability proportional to $\|\vec n_x\|( \min_{j \in T} \| \hat n_x - \hat n_j\|^2)$ and add the index $x$ to the list $T$. 
		\item Repeat step $(b)$ until we have an initial set of $r$ distinct vectors.
	\end{enumerate}
	\item Compute $c^* = \arg \min_{1\leq c \leq r} \| \vec n_x- \vec u_c\|^2$ for each $x$, and assign the group $\vec n_x \in G_{c^*}$.
	\item Update $\vec u_c$ such that it points along the sum of the elements of its associated group
	\begin{equation}
		\vec u_c \leftarrow \frac{\sum_{x \in G_c} \vec n_x}{\left \| \sum_{x \in G_c} \vec n_x \right \|}.
	\end{equation}
	\item Repeat step $2-3$ until the assignment stops changing.
\end{enumerate}
Typically, we run this clustering algorithm a few dozen times.

\subsubsection{Data adapted candidates 2}
When $r$ is comparable with $n$, the clustering method described above degenerates, as almost every group $G_c$ contains a single vector. Thus, in the regime $r \sim n$ we instead use a hierarchical clustering method \cite{agglo_clustering}. The main idea of this algorithm is to group together the preparations $\vec n_x$ into clusters based on their proximity. 
\begin{enumerate}
	\item To each vector $\vec n_x$, assign a group $G_x$ with $x = 1,\dots,n$
	\item Identify which two groups $G_i$ and $G_j$ are closest based on their Euclidean distance: $\min_{i\neq j} \|G_i\|+\|G_j\| - \|G_i+G_j\|$ with $\| G_i\|=\|\sum_{x\in G_i} \vec{n}_x\|$. Combine these two groups into a new group.
	\item Recalculate the distance between the newly formed cluster and all other clusters and merge the two closets groups.
	\item Repeat step 3 until $r$ clusters are formed.
	\item Assign the $r$ Bloch vectors in $\mathcal U$ as follows
	\begin{equation}
		\vec u_c \leftarrow \frac{\sum_{x \in G_c} \vec n_x}{\left \| \sum_{x \in G_c} \vec n_x \right \|}.
	\end{equation}
\end{enumerate}

\subsubsection{Jittered candidates}
Lastly, we add "jittered" candidates to fill out the pool $\mathcal A$. In this case we draw a random vector $\vec u$, and for each $c = 1,\dots,r$ we add random noise $\vec g_c$ of size $0.15$. That is, 
\begin{equation}
	\vec u_{c} \leftarrow \frac{\vec u + 0.15 \ \vec{g}_c}{\|\vec u + 0.15 \vec{g}_c\|}
\end{equation}

\subsection{Outer approximation: constructing the witness}
To obtain a good witness to detect $r$-complexity we want to perform the following min-max optimization 
\begin{equation}\label{eq:witness_optimization}
	\begin{aligned}
		&H = \min_{\vec w_1,\dots\vec w_n } h(w)  \\
		&\text{s.t.} \quad h(w) = \max_{\vec u_1,\dots,\vec u_r}  \sum_{x=1}^n \max_c \langle \vec w_x, \vec u_c\rangle \\
		& \qquad \sum_{x=1}^n\langle \vec w_x,\vec n_x\rangle = 1.
	\end{aligned}
\end{equation}
While this min-max optimisation problem is difficult to solve in practice, we perform an alternating search over $w = (\vec w_1,\dots \vec w_n)$ and $u = (\vec u_1,\dots \vec u_r)$ such that the normalisation constraint $\langle w, \mathcal E\rangle = 1$ remains satisfied. Here we have used the notation $\langle w, \mathcal E\rangle =  \sum_x \langle \vec w_x, \vec n_x\rangle$.

\begin{enumerate}
	\item Initialise $\vec w_x = \vec n_x / \|\mathcal E\|^2$ where $\|\mathcal E\| = \left(\sum_{x = 1}^n \|n_x\|^2\right)^{1/2}$. This choice of $w$ automatically satisfies the normalisation constraint $\langle w, \mathcal E\rangle = 1$.
	\item Fix the value of $w$. Depending on the ratio $r/n$ we then employ one of the two clustering methods described in the sections \textit{Data adapted candidates} to search for a suitable tuple $u$ with respect to $w$. This returns the witness $h(w)$. 
	\item Fix $u$, and minimise  the convex function $h(w)$ over $w$ using a subgradient method \cite{Subgradient_2014}. For completeness we summarise the subgradient method below.
	\begin{enumerate}
		\item Compute the subgradient $g = \partial_w h$. Each component $g = (\vec g_1,\dots,\vec g_n)$ reads $\vec g_x = \vec u_{c^*(x)}$ with $c^*(x)= \arg \max_{c} \langle \vec w_x,\vec u_c \rangle$.
		\item Remove the component of $g$ pointing along $\mathcal E = \{\vec n_x\}_x$. This is done using a standard orthogonal projection
		\begin{equation}
			g \leftarrow g - \frac{\langle g, \mathcal E \rangle}{\|\mathcal E\|^2}\mathcal E.
		\end{equation}
		Here $\langle g, \mathcal E\rangle = \sum_x \langle \vec g_x, \vec n_x\rangle$ and $\|\mathcal E\|^2 = \sum_x \|\vec n_x\|^2$. This step ensures that  $\langle w, \mathcal E\rangle = 1$ remains satisfied when $w$ is updated in the direction $-g$.
		\item Next, compute the step length $\alpha_k = \theta_k (H_{\text{best}}^{(k)} - v_{\text{lb}})/\|g\|_2^2$. Here, $v_{\text{lb}}$ is the lower bound on the visibility obtained from the interior point search in the previous section and $\theta_k = 1/(1+k/300)$. The factor $\theta_k$ results in that the subgradient method makes larger step sizes early and increasingly smaller corrections later.
		\item Update the witness
		\begin{equation}
			w \leftarrow w - \alpha_k g
		\end{equation} 
		and then restore normalisation to ensure that numerical errors do not propagate through the iterations
		\begin{equation}
			w \leftarrow w+ \frac{1-\langle w, \mathcal E\rangle}{\| \mathcal E\|^2} \mathcal E.
		\end{equation}
		
	\end{enumerate}
	\item Repeat step $2-3$ a few hundred times. At each iteration $k$ we keep $H_{\text{best}}^{(k)} = \min \{H_{\text{best}}^{(k-1)}, h(w)^{(k)}\}$
\end{enumerate}
Since search for a good set $\vec u$ using a clustering method instead of performing a faithful optimisation over all possible sets of Bloch vectors, we do not reach the optimal value of $h(w)$. Thus, this procedure returns a good witness candidate $w$, but the value $h(w)$ does not correspond to a certified upper bound on the critical visibility.

\subsubsection{Subgradient method}
This summary is based on Ref~\cite{Subgradient_2014}. The subgradient method is a simple algorithm to minimise a convex function. Unlike the standard gradient method, the subgradient method is directly applicable to non-differentiable functions and its step lengths are fixed ahead of time. However, the subgradient method is not a descent method, meaning that the function value is not guaranteed to decrease at all steps.

Consider the convex function $f(x): \mathbb R^n \rightarrow \mathbb R$. The goal is to minimise $f$ over $x$. To do this, the subgradient method uses the iteration
\begin{equation}
	x^{(k+1)} = x^{(k)} - \alpha_k g^{(k)}.
\end{equation}
Here, $x^{(k)}$ is the $k$th iterate, $g^{(k)} \in \partial f(x^{(k)})$ is the subgradient of $f$ at $x^{(k)}$, and $\alpha_k > 0$ is the $k$th step size. That is, at each iteration we take a step in the direction of the negative subgradient. Since $g^{(k)}$ might not be a descent direction, we keep track of the best point found so far at each step, i.e., we set 
\begin{equation}
	f_{\text{best}}^{(k)} = \min \{f_{\text{best}}^{(k-1)}, f(x^{(k)})\},
\end{equation}
and if $f_{\text{best}}^{(k)}$ is the best point so far we store the index of the associated step $i_{\text{best}} = k$. When the optimal value $f^*$ of the function is known,  it is a good choice to use a subgradient step length due to Polyak, namely
\begin{equation}\label{eq:polyak_ideal}
	\alpha_k = \frac{f(x^{(k)})-f^*}{\|g^{(k)}\|^2}.
\end{equation}
To motivate this step,  assume that $\alpha$ is small and that $g = \nabla f(x)$. Using Taylor expansion, up to first order in $\alpha$, we then find that
\begin{equation}
	f(x-\alpha g) = f(x) - g^T(x - (x-\alpha g)) = f(x) - \alpha \|g\|^2
\end{equation}
Replacing the left-hand side by $f^*$ and solving for $\alpha$ yields Eq~\eqref{eq:polyak_ideal}. If the optimal value of the function is not known but we have a genuine lower bound $f_{\text{lb}} \leq f^*$ a good practical choice is
\begin{equation}
	\alpha_k = \theta_k \frac{f(x^{(k)})-f_{\text{lb}}}{\|g^{(k)}\|^2},
\end{equation}
where $0 \leq \theta_k \leq 1$ is used to adapt the step lengths.

\subsection{Implementation of certificate A}
We now describe how to implement certificate A in practice. That is, we compute the upper bound on $h_Y(w)$ using the expression $\hat h_Y(w)$ in Observation ~\ref{result:L_certificate}. Since $\hat h_Y$ is a non-differentiable convex function we use the subgradient method to perform the minimisation over $\mu = (\mu_1,\dots,\mu_n)$. To this end we implement the following algorithm

\begin{enumerate}
	\item Generate an origin-centred polytope approximation $Y$ of the Bloch sphere with $N$ points with associated shrinking factor $s > 0$. We do this using the package \textit{EntanglementDetection.jl} introduced in Ref~\cite{Liu_2025}. 
	\item  Construct an initial set $\mu$  and define $L \equiv \sum_x \mu_x$. We describe how the initial set $\mu$ is generated below.
	\item Compute all $G_j(\mu) = \sum_x  \max(g_x(j)-\mu_x,0)$ for all $j= 1,\dots,N$ with $g_x(j) = \langle \vec w_x, \vec y_j\rangle$. We then sort them, collect the $r$ largest $G_j(\mu)$ and store the associated indices $j$ in a list $T$.
	\item Compute $\hat h_Y(\mu) = \sum_x \mu_x + \sum_{j \in T} G_j(\mu)$. 
	\item Compute the subgradient $d\mu = \partial_{\mu} \hat h_Y$, where each component reads
	\begin{equation}
		d\mu_x = 1- \sum_{j \in T } \theta_{xj}, \qquad \theta_{xj} = 
		\begin{cases} 1, \quad  g_x(j) > \mu_x \\  
			0, \quad  g_x(j) \leq \mu_x
		\end{cases}.
	\end{equation}
	\item We then update the variables $\mu$ and take a step in the direction of the negative gradient
	\begin{equation}
		\mu \leftarrow \max(0, \mu - \alpha_k d\mu)
	\end{equation}
	where $\alpha_k =g_L \theta /\|d\mu\|^2$ with $g_L = \max(\hat h-L,0)$ and $\theta = 1.5 / (1 + k/60)$ where $k$ counts the iteration.
	\item Repeat step $3-6$ a few hundred times, i.e., $k \simeq 0,\dots,500$.  At each iteration we keep $\hat h_{\text{best}} \leftarrow \min \{\hat h_{\text{best}}, \hat h_Y\}$ and $\hat h_Y$ is the value attained at the $k$th step.
	\item Return $\hat h_Y = \hat h_{\text{best}}/s$. 
\end{enumerate}

\hfill

Lastly, we explain how to construct the initial set $\mu$. This is done by finding a good set $T = \{\vec y_j\}_{j}$ of $r$ Bloch vectors $\vec y_j \in Y$ of the polytope approximation, such that $S = \sum_x \max_{j\in T} g_x(j)$ is large where $g_x(j) = \langle \vec w_x, \vec y_j\rangle$. This is done following the algorithm

\begin{enumerate}
	\item Introduce variables $\text{current}_x = -\infty$ for $x = 1,\dots,n$, and initialise the empty list $T$
	\item Introduce variables $S_{\text{best}} = -\infty$ and $i_{\text{best}}$. 
	\item For each point $\vec y_j$ with $j \notin T$ evaluate the expression
	\begin{equation}
		S_j = \sum_x \max(g_x(j), \text{current}_x).
	\end{equation}
	If $S_j > S_{\text{best}}$ update $S_{\text{best}} \leftarrow S_j$ and $i_{\text{best}} \leftarrow j$.
	\item Add $i_{\text{best}}$ to the list $T$ and take $\text{current}_x \leftarrow \max(g_x(i_{\text{best}}),\text{current}_x)$.
	\item Repeat step $2-4$ $r$ times.
\end{enumerate}
After we have found $r$ good points $\vec{y}_j$, we let $\mu_x = \max_{j \in T} g_x(j)$ and define $L \equiv \sum_x \mu_x$.

.

\subsection{Case studies}
In this section we analyse the performance of the algorithms introduced above. We do three types of case studies. We first take $\mathcal E$ to form a Fibonacci lattice of $n$ points. Thereafter, we select $\mathcal E$ to be a set of $n$ Haar-random pure states using the function \textit{random\_state\_ket} from the package Ket.jl \cite{ket_2026}. Lastly, we randomly generate mixed states using the function \textit{random\_state} from the package Ket.jl \cite{ket_2026}, which uniformly samples a state according to the Hilbert-Schmidt metric. For such sampling of $\mathcal E$ we then evaluate the lower and upper bounds on the critical visibility using the inner and outer approximations of $r$-complexity for different $n$ and $r$. The bounds are summarised in Tables \ref{tab:near-uniform}, \ref{tab:random-pure}, and  \ref{tab:random-mixed}. We further include the time it took to run the algorithms on a standard laptop. We see that the algorithms are both efficient and accurate. Moreover, we observe that the critical visibility decreases when the ratio $r/n$ decreases, whereas it is almost unit when $r\sim n$, as expected. By comparing the table for the pure states placed in a Fibonacci lattice and pure Haar random states, we note that the visibilities are similar. This is expected, since both methods approximate a uniform distribution of qubit states for large enough $n$. For these types of ensembles the difference between the upper and lower bounds is typically within one percentage point. In contrast, the gap between the bounds increases when we apply the search algorithms to mixed states. For the random mixed ensembles $\mathcal E$ we further note that an increase in the message size $r$ also yields a higher performance of the CCMs compared to the case in which the ensemble only contains pure states. This can be understood by the fact that mixed states already are a convex combination of pure states.

\begin{table}[htbp]
	\centering
	\caption{Results for sets of $n$ Bloch vectors on a Fibonacci lattice \cite{Fibonacci_lattice} for different $r$. The time $t$ needed to compute the lower $v_{\text{lb}}$ and upper $v_{\text{ub}}$ bounds on the critical visibility is given in seconds.}
	\label{tab:near-uniform}
	\setlength{\tabcolsep}{8pt}
	\begin{tabular}{cc|cccc}
		\toprule
		$n$ & $r$ & \textbf{Lower bound} & \textbf{Upper bound} & \textbf{Time lower (s)} & \textbf{Time upper (s)} \\
		\midrule
		100 & 2   & 0.5007 & 0.5033 & 23  & 2   \\
		100 & 3   & 0.6506 & 0.6714 & 15  & 3   \\
		100 & 4   & 0.7459 & 0.7549 & 16  & 5   \\
		100 & 8   & 0.8704 & 0.8811 & 30  & 11  \\
		\addlinespace
		200 & 2   & 0.4999 & 0.5027 & 157 & 4   \\
		200 & 3   & 0.6496 & 0.6700 & 29  & 7   \\
		200 & 4   & 0.7450 & 0.7535 & 57  & 9   \\
		200 & 8   & 0.8690 & 0.8787 & 114 & 22  \\
		\addlinespace
		400 & 2   & 0.4993 & 0.5022 & 131 & 9   \\
		400 & 3   & 0.6493 & 0.6697 & 70  & 14  \\
		400 & 4   & 0.7447 & 0.7530 & 360 & 21  \\
		400 & 8   & 0.8672 & 0.8780 & 594 & 52  \\
		\addlinespace
		800 & 2   & 0.4988 & 0.5021 & 482 & 16  \\
		800 & 3   & 0.6491 & 0.6696 & 212 & 28  \\
		800 & 4   & 0.7445 & 0.7528 & 247 & 59  \\
		800 & 8   & 0.8661 & 0.8777 & 875 & 195 \\
		\addlinespace
		1000 & 2  & 0.4970 & 0.5061 & 456 & 4   \\
		1000 & 3  & 0.6480 & 0.6726 & 206 & 6   \\
		1000 & 4  & 0.7436 & 0.7555 & 236 & 8   \\
		1000 & 8  & 0.8653 & 0.8809 & 972 & 12  \\
		\addlinespace
		2000 & 2  & 0.4961 & 0.5068 & 1379 & 10 \\
		2000 & 3  & 0.6487 & 0.6723 & 1512 & 13 \\
		2000 & 4  & 0.7441 & 0.7553 & 1124 & 15 \\
		2000 & 8  & 0.8638 & 0.8827 & 900  & 25 \\
		\addlinespace
		5000 & 2  & 0.4960 & 0.5080 & 8702 & 26 \\
		5000 & 3  & 0.6486 & 0.6719 & 7579 & 30 \\
		5000 & 4  & 0.7440 & 0.7553 & 9032 & 36 \\
		5000 & 8  & 0.8628 & 0.8814 & 4844 & 60 \\
		\addlinespace
		10000 & 2 & 0.4712 & 0.5375 & 9156  & 50  \\
		10000 & 3 & 0.6420 & 0.6737 & 6158  & 88  \\
		10000 & 4 & 0.7400 & 0.7581 & 8031  & 104 \\
		10000 & 8 & 0.8617 & 0.8824 & 17583 & 180 \\
		\addlinespace
		150 & 10  & 0.8948 & 0.9061 & 186 & 19  \\
		150 & 40  & 0.9741 & 0.9816 & 256 & 230 \\
		150 & 75  & 0.9860 & 0.9927 & 356 & 380 \\
		150 & 110 & 0.9912 & 0.9976 & 436 & 21  \\
		150 & 140 & 0.9941 & 0.9994 & 549 & 7   \\
		150 & 149 & 0.9957 & 0.9999 & 364 & 4   \\
		\addlinespace
		300 & 75  & 0.9857 & 0.9918 & 280 & 382 \\
		300 & 150 & 0.9925 & 0.9977 & 171 & 145 \\
		300 & 225 & 0.9954 & 0.9988 & 261 & 163 \\
		300 & 290 & 0.9971 & 0.9999 & 360 & 48  \\
		300 & 299 & 0.9977 & 1.0000 & 320 & 32  \\
		\addlinespace
		500 & 200 & 0.9945 & 0.9966 & 1796 & 980 \\
		500 & 250 & 0.9956 & 0.9986 & 2638 & 930 \\
		500 & 400 & 0.9978 & 0.9994 & 2870 & 754 \\
		500 & 490 & 0.9996 & 0.9999 & 6288 & 162 \\
		500 & 499 & 0.9986 & 1.0000 & 3165 & 121 \\
		\bottomrule
	\end{tabular}
\end{table}

\begin{table}[htbp]
	\centering
	\caption{Results for the $n$ random pure states using an $r$-valued message. The time $t$ needed to compute the lower $v_{\text{lb}}$ and upper $v_{\text{ub}}$ bounds on the critical visibility is given in seconds.}
	\label{tab:random-pure}
	\setlength{\tabcolsep}{8pt}
	\begin{tabular}{cc|cccc}
		\toprule
		$n$ & $r$ & \textbf{Lower bound}  & \textbf{Upper bound}  & \textbf{Time lower (s)}& \textbf{Time upper (s)} \\
		\midrule
		100 & 2   & 0.5016 & 0.5059 & 57  & 2  \\
		100 & 4   & 0.7475 & 0.7608 & 25  & 6  \\
		100 & 8   & 0.8688 & 0.8865 & 29  & 9  \\
		\addlinespace
		300 & 2   & 0.4995 & 0.5039 & 160 & 9  \\
		300 & 4   & 0.7451 & 0.7563 & 126 & 15 \\
		300 & 8   & 0.8662 & 0.8809 & 242 & 39 \\
		\addlinespace
		60  & 15  & 0.9364 & 0.9560 & 9  & 20 \\
		60  & 45  & 0.9874 & 0.9980 & 16 & 4  \\
		60  & 59  & 0.9975 & 0.9999 & 13 & 4  \\
		\addlinespace
		120 & 60  & 0.9858 & 0.9955 & 31 & 27 \\
		120 & 90  & 0.9937 & 1.0000 & 48 & -- \\
		120 & 119 & 0.9997 & 1.0000 & 48 & -- \\
		\bottomrule
	\end{tabular}
\end{table}

\begin{table}[htbp]
	\centering
	\caption{Results for the $n$ random mixed states for an $r$-valued message. The time $t$ needed to compute the lower $v_{\text{lb}}$ and upper $v_{\text{ub}}$ bounds on the critical visibility is given in seconds. }
	\label{tab:random-mixed}
	\setlength{\tabcolsep}{8pt}
	\begin{tabular}{rrcccc}
		\toprule
		$n$ & $r$ &  $v_{\text{lb}}$  & $v_{\text{ub}}$  & $t_{\text{lb}} (s)$ & $t_{\text{ub}} (s)$\\
		\midrule
		100 & 2 & 0.5377 & 0.5623 & 8  & 2  \\
		100 & 4 & 0.8004 & 0.8385 & 6  & 6  \\
		100 & 8 & 0.9263 & 0.9741 & 12 & 12 \\
		\addlinespace
		300 & 2 & 0.5144 & 0.5463 & 60 & 10 \\
		300 & 4 & 0.7697 & 0.8046 & 54 & 21 \\
		300 & 8 & 0.8974 & 0.9348 & 86 & 50 \\
		\bottomrule
	\end{tabular}
\end{table}

\end{document}